\newif\ifsubmode
\submodefalse

\newif\ifprintfig
\printfigtrue

\newif\ifemulate
\emulatetrue

\ifsubmode
  \documentstyle[12pt,aasms4,epsf]{article}
  \received{}
  \accepted{}
  \journalid{}{}
  \articleid{}{}
\else
   \ifemulate
     \documentstyle[emulateapj,epsf,graphicx]{article}
     \submitted{{\it Accepted for publication in ApJ}}
   \else
     \documentstyle[11pt,aaspp4,epsf]{article} 
     \slugcomment{{\it Accepted for publication in ApJ}}
   \fi
\fi

\lefthead{Frank C. van den Bosch}
\righthead{Semi-Analytical Models for the Formation of Disk Galaxies}

\newcommand{\etal}{{et al.~}}

\newcommand{\lta}{\lesssim}
\newcommand{\gta}{\gtrsim}

\newcommand{\lte}{\leq}
\newcommand{\gte}{\geq}

\newcommand{\kmsmpc}{\>{\rm km}\,{\rm s}^{-1}\,{\rm Mpc}^{-1}}
\newcommand{\kms}{\>{\rm km}\,{\rm s}^{-1}}

\newcommand{\Msun}{\>{\rm M_{\odot}}}
\newcommand{\Lsun}{\>{\rm L_{\odot}}}

\begin{document}

\title{Semi-Analytical Models for the Formation of Disk Galaxies: \\
       I. Constraints from the Tully-Fisher Relation}

\ifemulate
  \author{Frank C. van den Bosch\altaffilmark{1}}
  \affil{Department of Astronomy, University of Washington, Seattle, 
         WA 98195, USA  vdbosch@astro.washington.edu}
\else
  \author{Frank C. van den Bosch\altaffilmark{1,2}}
  \affil{Department of Astronomy, University of Washington, Seattle, 
         WA 98195, USA}
\fi


\altaffiltext{1}{Hubble Fellow}

\ifemulate\else
  \altaffiltext{2}{{\tt vdbosch@astro.washington.edu}}
\fi


\ifsubmode\else
  \ifemulate\else
     \clearpage
  \fi
\fi


\ifsubmode\else
  \ifemulate\else
     \baselineskip=14pt
  \fi
\fi


\begin{abstract}
  We present new  semi-analytical  models for  the formation of   disk
  galaxies with  the purpose   of   investigating the origin   of  the
  near-infrared Tully-Fisher  (TF) relation.   The models assume  that
  disks are formed by  cooling of the  baryons inside dark halos  with
  realistic density  profiles,  and  that the  baryons  conserve their
  specific angular momentum.   Adiabatic contraction of  the dark halo
  is taken into account, as well as a recipe for bulge formation based
  on a self-regulating mechanism that ensures disks to be stable. Only
  gas  with  densities above the critical    density given by Toomre's
  stability criterion is considered   eligible for star formation.   A
  Schmidt law is  assumed to prescribe the  rate at which this gas  is
  transformed  into stars.   The   introduction of the  star formation
  threshold density proves an essential ingredient  of our models, and
  yields    gas mass fractions  that  are  in excellent agreement with
  observations.  Finally,  a simple recipe  for supernovae feedback is
  included.   We emphasize the   importance  of extracting the  proper
  luminosity and velocity measures from the models, something that has
  often been  ignored in  the  past.  We  use   the zero-point of  the
  $K$-band TF relation  to place stringent constraints on cosmological
  parameters.  In particular, we rule out  a standard cold dark matter
  universe, in which disk galaxies are too faint to be consistent with
  observations.     The  TF     zero-point,   in  combination     with
  nucleosynthesis    constraints  on the  baryon   density,   and with
  constraints  on the normalization of the  power spectrum, requires a
  matter  density $\Omega_0   \lta  0.3$.  The   observed  $K$-band TF
  relation has  a slope that  is steeper than simple predictions based
  on dynamical arguments suggest.   Taking the stability  related star
  formation threshold densities into  account steepens the TF relation
  and decreases its scatter.  However, in order for the slope to be as
  steep as observed, further physics are  required.  We argue that the
  characteristics  of the  observed near-infrared  TF  relation do not
  reflect     systematic   variations   in    stellar  populations, or
  cosmological  initial    conditions.  In  fact,  feedback   seems an
  essential ingredient  in order to explain  the observed slope of the
  $K$-band TF relation.  Finally we  show  that our models  provide  a
  natural explanation for the small  amount of scatter that makes  the
  TF relation useful as a cosmological distance indicator.
\end{abstract}



\ifemulate\else
   \clearpage
\fi


\section{Introduction}
\label{sec:intro}

Within the standard picture   for galaxy formation, in which  galaxies
are  thought to  form out  of  the  gas which dissipatively  collapses
within the potential wells provided by virialized dark halos (White \&
Rees 1978),  disk galaxies take a  special  place.  Their flatness and
rotational   support suggest a   relatively  smooth  formation history
without violent non-linear processes.   The structure and  dynamics of
disk galaxies    are thus expected    to be  strongly related  to  the
properties  of the  dark   halos in  which  they  are  embedded.  This
principle is in  fact the key idea  behind the standard model for disk
formation which was set  out by Fall \&  Efstathiou (1980),  and which
has  since been extended upon  by  numerous investigators (Faber 1982;
Fall 1983; van der Kruit 1987; Dalcanton, Spergel \& Summers 1997; Mo,
Mao \&  White 1998; van  den Bosch 1998).   In  this picture,  the gas
radiates  its binding energy  but  retains its angular momentum,  thus
settling into a rotationally supported disk, the scale length of which
is proportional to both the size and the angular  momentum of the dark
halo.

Understanding galaxy  formation is intimately  linked to understanding
the origin of the fundamental scaling  relations of galaxies.  In this
paper we  use new semi-analytical  models  for the formation  of  disk
galaxies to investigate the origin of  the Tully-Fisher (hereafter TF)
relation (Tully \& Fisher 1977), which has the form
\begin{equation}
\label{TFpower}
L=A \, V_{\rm rot}^{\gamma},
\end{equation}
or, in terms of a linear law in a log-log plane:
\begin{equation}
\label{TFlinear}
M = a + b \, {\rm log} V_{\rm rot},
\end{equation}
where $M$ is the absolute magnitude of the disk, and $b = -2.5 \gamma$
is called  the  slope of the relation.    In the light   of the strong
coupling  between disk   and dark  halo   mentioned above,  a  scaling
relation like  this  is  central to  theories   of galaxy  formation.  
Understanding its origin   can put constraints  on theories  of galaxy
formation,  and in particular on the  detailed  physics that cause the
baryonic  component of a dark halo  to be transformed  into a luminous
galaxy.   As such,  any successful  theory  of (disk) galaxy formation
should be able  to explain the slope, zero-point,  and small amount of
scatter of this fundamental scaling relation.

Numerous  studies in  the past  have addressed  the origin  of  the TF
relation.  However, no consensus has been reached, and it is currently
still under debate whether  the  origin of  the TF relation  is mainly
governed by initial cosmological conditions  (e.g., Eisenstein \& Loeb
1996; Avila-Reese, Firmani \& Hern\'andez 1998; Firmani \& Avila-Reese
1998a,b), or by the detailed processes  governing star formation (Silk
1997; Heavens \& Jiminez 1999) and/or feedback (e.g., Kauffmann, White
\&  Guiderdoni 1993; Cole \etal 1994;   Elizondo \etal 1999; Natarajan
1999).  

One of the main reasons for this lack of consistency is the discordant
use  of luminosity and rotation  measures  in TF relations.  The slope
and scatter of the TF relation depend strongly on the photometric band
used  to measure  the luminosities: values   of $\gamma$ increase from
$\sim   2.5$  in the  $B$-band   to $\sim 4$    in the infrared (e.g.,
Aaronson,   Huchra \& Mould  1979; Visvanathan   1981; Tully, Mould \&
Aaronson 1982; Wyse  1982; Pierce    \&  Tully 1988; Gavazzi     1993;
Verheijen    1997).   For the   rotation  velocity  a  large number of
different measures  have  been used: HI  line widths  from single-dish
observations,  velocities  measured  from  H$\alpha$ rotation  curves,
rotation velocities at a  fixed number of  disk scale lengths, maximum
observed rotation velocities, rotation velocities at the last measured
point, and the velocity of the  flat part of the  rotation curve.  All
these  different  measures yield  wildly   different TF relations (see
Courteau 1997 and Verheijen 1997  for detailed discussions). Therefore
it is crucial that one extracts the same quantities from the models as
the  ones  used in  the empirical relation  whose  origin one seeks to
explain.  Most previous studies have ignored the subtleties associated
with these different measures,  and allowed themselves the ``freedom''
to compare  their model predictions with  the best fitting TF relation
available in the literature.

The other reason why we  still lack consensus on the  origin of the TF
relation regards the modeling techniques that have been used.  Several
investigators  have used numerical  simulations (e.g., Evrard, Summers
\&  Davis 1994;  Navarro \&  White  1994; Steinmetz \&  M\"uller 1994;
Tissera,  Lambas \&   Abadi 1997; Dom\'\i  nguez-Tenreiro,  Tissera \&
S\'aiz 1998; Steinmetz  \&  Navarro  1999; Elizondo  \etal   1999).  A
general problem,  however, is  that   the disks   that form in   these
simulations have specific angular momenta  that are more than an order
of magnitude lower than those of observed disks.  In addition, most of
these simulations ignore star  formation and feedback, making a direct
comparison   between   simulated  and observed  disks    a treacherous
enterprise (see  Evrard 1997  for a review).  Semi-analytical modeling
(SAM) of galaxy  formation has  been more  successful  (White \& Frenk
1991;  Cole 1991; Kauffmann,   White  \& Guiderdoni 1993; Lacey  \etal
1993; Cole \etal  1994; Heyl \etal 1995;  Somerville \& Primack 1998). 
Despite reasonable success  in reproducing certain TF relations, these
models are hampered  by    several shortcomings (see    discussion  in
\S~\ref{sec:model}).   More   recently,  Avila-Reese,  Firmani      \&
Hern\'andez  (1998; hereafter    AFH98)  and Firmani  \&   Avila-Reese
(1998a,b; hereafter  FA98) presented  new SAMs  that focus only on
the formation  of disk galaxies.   These models improve upon  previous
SAMs in several  important   ways.  From a detailed   comparison  with
observed   TF relations, these authors   conclude that the TF relation
represents a fossil of the primordial density fluctuation field.

In this paper we present  new semi-analytical models for the formation
of  disk galaxies. We improve  upon previous studies  by  (i) making a
well-motivated  choice for the   luminosity and velocity measures that
define  the TF relation  whose origin we  seek  to understand, (ii) by
extracting the same  measures from the  models, and (iii) by using new
semi-analytical models that focus  on  the formation of disk  galaxies
and that  alleviate several shortcomings  of previous models. The main
aim  of this paper is to  investigate the prevailing mechanism that is
responsible for the  slope, zero-point and small  amount of scatter of
the TF relation. Although our  models are, in many ways, complementary
to those of FA98, we reach a different conclusion.

\section{The Empirical Tully-Fisher Relation}
\label{sec:obs}

One  of the most detailed investigations  of the empirical TF relation
is  that of  Verheijen (1997, hereafter  V97), who  obtained $B$, $R$,
$I$, and    $K$ band photometry, as   well   as detailed  HI synthesis
observations of a complete  sample of (bright)  spiral galaxies in the
Ursa Major cluster (see  also Tully \etal 1996).   V97 made a detailed
comparison of different TF relations using different photometric bands
and different velocity  measures.  This study  has yielded a number of
important results: (i) the scatter in the TF relation is smallest when
using  the  flat part  of the  HI rotation curve,   $V_{\rm flat}$, as
velocity measure, (ii) when using the HI line widths, the TF relations
become shallower and harbor  a  larger scatter,  (iii) the  results of
Sprayberry  \etal (1995)  and  Zwaan \etal (1995),  that high  surface
brightness  (HSB) and low surface  brightness (LSB) spirals follow the
same TF relation, are  confirmed, (iv) both  the scatter and the slope
of the TF relation are extremely  sensitive to selection criteria, and
(v) the slope of the TF relation changes from $b=-6.7$ in the $B$-band
to $-10.5$ in the $K$-band.

Luminosities depend more  strongly on extinction, stellar populations,
star formation   histories, and   metallicities going    towards bluer
passbands.   Given  the  large  uncertainties associated with  stellar
population models  and with a  proper treatment of extinction by dust,
and since the  TF relation is  fundamentally a dynamical law, the most
natural passbands to  use are   those in  the near-infrared in   which
luminosity is most directly  related to the  luminous mass  (see e.g.,
Gavazzi 1993; Gavazzi, Pierini \& Boselli  1996).  If we further limit
ourselves to TF relations that are not based  on HI line widths, since
these  are   particularly hard to   interpret  in terms of fundamental
parameters of the galaxy (e.g., Haynes \etal 1999), the available data
becomes actually very limited.  The only TF  relation that is based on
full near-infrared imaging (rather  than aperture photometry) and  has
velocity measures  that are derived from full   HI rotation curves, is
the $K$-band TF relation of V97:
\begin{equation}
\label{TFfund}
M_K - 5 \, {\rm log} h = 0.683 - 10.5 \, {\rm log} V_{\rm flat},
\end{equation}
Here $V_{\rm flat}$ is the measured rotation velocity at the flat part
of the  HI rotation curve  and $h  =  H_0/(100 \kms {\rm  Mpc}^{-1})$. 
Magnitudes have been converted to $M_K - 5  \, {\rm log} h$ adopting a
distance to the  Ursa Major cluster  of 15.5 Mpc, which corresponds to
$H_0 = 85 \kmsmpc$  (Pierce \& Tully 1988, 1992).  In what follows  we
consider  equation~(\ref{TFfund}) the  `fundamental' TF relation whose
origin we seek to understand.

\vskip 3.0truecm

\section{Modeling the Formation of Disk Galaxies}
\label{sec:model}

In  order to investigate   whether the origin   of the TF relation  is
mainly  related  to  cosmological  initial conditions,  or  to details
regarding the  processes   associated  with   star formation    and/or
supernovae feedback,  we  construct  new semi-analytical  models  that
focus on the  near-infrared  properties  of disk  galaxies only.    We
incorporate a number of improvements over previous  SAMs, all of which
(except  for FA98) assumed  all dark halos  to have  the same specific
angular momentum.  Since the angular momentum sets, to a large degree,
the surface density of the disk, and since star  formation is a strong
function of surface density  (see \S~\ref{sec:sf}), a realistic spread
in halo specific  angular momenta yields  a spread in luminosities for
galaxies forming  in  halos of   the  same mass.    This is especially
important  when investigating for   example   the scatter in  the   TF
relation.   Also,   we  take    adiabatic contraction    into  account
(Blumenthal  \etal 1986;  Flores  \etal  1993)  and consider the  more
realistic  universal   halo density  profiles,  rather  than  singular
isothermal spheres,  which have been   found to poorly  fit the actual
density  profiles of dark halos in   $N$-body simulations (e.g., Frenk
\etal 1988; Efstathiou \etal 1988; Dubinski \& Carlberg 1991; Navarro,
Frenk \& White 1995,  1996, 1997; Tormen, Bouchet  \& White 1997).  We
use a star formation recipe that is  different from previous SAMs, and
mimic observations of the actual  rotation curve of the galaxy, rather
than to simply assume that the observed rotation  velocity is equal to
the circular velocity at the halo's  virial radius.  As we show below,
these latter two modifications prove to be of crucial importance.
 
The dynamical  fragility of disks implies  that their mass aggregation
histories can   not have been too violent   (e.g., T\'oth  \& Ostriker
1992).  We   therefore   refrain  from  calculating   complete  merger
histories of  the dark halos.  The study  by  Navarro \etal (1997) has
shown  that virialized dark  halos  are well described  by a universal
density profile,  whose    properties can  be   calculated  from   the
Press-Schechter  formalism.  In a recent study,  Jing (1999) has shown
that    halos with significant    amounts of  substructure can deviate
significantly from   this universal  density  profile.  However, these
halos are  unlikely to yield disk  galaxies, and the universal density
profiles can thus   be used to  describe   the end-state  of  the mass
aggregation histories of halos  surrounding spiral galaxies  (see also
AFH98).

As shown by Baugh,  Cole \& Frenk   (1996), the merging  histories and
star formation  histories are well  coupled.  Ignoring (small) mergers
therefore mainly reflects on  the details regarding the star formation
histories. Since  we focus  on the  near-infrared  properties,  we are
insensitive to such details, given the  fact that mass-to-light ratios
in the near-infrared depend only very weakly on age and/or metallicity
(see  \S~\ref{sec:stelpop}   below).    Furthermore,  the  quiet  mass
aggregation histories of disk  galaxies implies smooth  star formation
histories  with little scatter (Baugh \etal  1996).  Therefore, in the
SAMs presented here we do not follow  the detailed {\it time-resolved}
formation of disk galaxies,  but limit ourselves  to modeling the {\it
  time-integrated} properties.

\subsection{Properties of the dark matter halos}
\label{sec:darkhalo}

Virialized dark   halos  can be   quantified by a  mass,  $M_{200}$, a
radius,  $r_{200}$,  and an angular  momentum,   $J$, originating from
cosmological torques (e.g., Hoyle 1953).  Here $r_{200}$ is defined as
the radius inside of  which the average  density of the system is  200
times the critical density  of the Universe,  and $M_{200}$  is simply
the total mass inside $r_{200}$.  In a series of papers Navarro, Frenk
\& White  (1995, 1996,  1997) showed that  virialized dark  halos have
universal  equilibrium  density profiles,    independent  of mass   or
cosmology, which can be well fit by
\begin{equation}
\label{nfwprof}
\rho(r) = \rho_{\rm crit} {\delta_0 \over (r/r_s) (1 + r/r_s)^2},
\end{equation}
where
\begin{equation}
\label{overdens}
\delta_0 = {200 \over 3} {c^3 \over {\rm ln}(1+c) - c/(1+c)},
\end{equation}
with $c=r_{200}/r_s$ the  concentration  parameter, and $r_s$ a  scale
radius.   Given the mass  $M_{200}$  and  the specific cosmology,  the
concentration  parameter  $c$  can be  calculated using  the procedure
outlined in  the Appendix of Navarro  \etal (1997). In what follows we
only consider spherical halos, and we  refer to halos with the density
profile of equation~(\ref{nfwprof}) as NFW halos.

The angular  momentum of the  halo is quantified  by the dimensionless
spin parameter, $\lambda$, defined by
\begin{equation}
\label{spinparam}
\lambda = {J \vert E \vert^{1/2} \over G M_{200}^{5/2}}.
\end{equation}
Here $E$ is the halo's  total energy. Several studies, both analytical
and  numerical, have  shown  that  the  distribution  of spin  angular
momenta   of collapsed dark  matter  halos is  well  approximated by a
log-normal distribution:
\begin{equation}
\label{spindistr}
p(\lambda){\rm d} \lambda = {1 \over \sigma_{\lambda} \sqrt{2 \pi}}
\exp\biggl(- {{\rm ln}^2(\lambda/\bar{\lambda}) \over 2
  \sigma^2_{\lambda}}\biggr) {{\rm d} \lambda \over
  \lambda},
\end{equation}
(e.g.,  Barnes \& Efstathiou  1987; Ryden  1988; Cole  \& Lacey  1996;
Warren \etal 1992).  We   consider a  distribution that  peaks  around
$\bar{\lambda} = 0.05$ with $\sigma_{\lambda} = 0.6$. These values are
in good agreement  with the $N$-body  results of  Warren \etal (1992),
and  virtually  independent of cosmology  (e.g.,   Lemson \& Kauffmann
1999).

\subsection{The formation of disk and bulge}
\label{sec:formation}

Once  the halo   properties  are known, we   can determine  the global
properties of  the disks that form  inside  them. The virialization of
the halo heats the  baryonic matter to  the virial temperature. As the
baryons start to cool they loose their  pressure support and settle in
a disk.   We  follow Fall \&  Efstathiou  (1980) and assume  that  the
specific  angular momentum of the  baryonic material is conserved.  We
further assume  that the disk  that  forms has  an exponential surface
density   profile. The scale-length  of  the  disk  is  proportional to
$\lambda$ and $r_{200}$, and  can be calculated using  equation~(28) in
Mo \etal (1998).

Self-gravitating   disks   tend   to   be   unstable   against  global
instabilities such as bar formation.   We use the stability  criterion
of Christodoulou, Shlosman \& Tohline (1995), according to which disks
are stable if they obey
\begin{equation}
\label{stabalpha}
\alpha = {V_d \over 2 V_c} < \alpha_{\rm crit},
\end{equation}
with  $V_d$ and $V_c$  the characteristic  circular velocities of  the
disk and the composite disk-bulge-halo system, respectively \footnote{
Throughout this paper we use the subscripts `$d$', `$b$', and `$h$' to
refer to  the disk, bulge,  and dark matter  halo, respectively.}.  As
characteristic velocities we consider the  circular velocities at $R =
3 R_d$, since Mo \etal (1998) have shown that this is the radius where
typically  the rotation  curve   of  the  composite system  reaches  a
maximum.   We set $\alpha_{\rm crit} =  0.35$ throughout, as suggested
for gaseous  disks (Christodoulou  \etal  1995).  We  have tested that
none  of  our  results  depend  significantly   on  the value  of this
parameter.

We follow the approach of van den Bosch (1998, 1999) and assume that a
gaseous   disk       which     is    unstable        according      to
criterion~(\ref{stabalpha}) transforms  part of its disk material into
a bulge component in a  self-regulating fashion,  such that the  final
disk is marginally  stable (i.e., $\alpha =  \alpha_{\rm crit}$).   We
describe the bulge by a sphere with a Hernquist density profile:
\begin{equation}
\label{bulgeprof}
\rho_b(r) = {M_b \over 2 \pi} \, {r_b \over r \, (r + r_b)^3},
\end{equation}
where $r_b$ is a scale length (Hernquist 1990). Andredakis, Peletier \&
Balcells (1995) have shown that the effective radius $r_e$ (defined as
the radius encircling half of the projected light) is directly related
to the total luminosity of the bulge by the empirical relation
\begin{equation}
\label{emprel}
M_B = -19.75 - 2.8 \, {\rm log}(r_e)
\end{equation}
If we  now use $M_B -  M_K = 4.0$,  in good agreement with the average
value for  bulges (Peletier \&  Balcells 1997), and   $r_e = 1.8153 \,
r_b$ (Hernquist 1990), we obtain
\begin{equation}
\label{bulgerad}
r_b = 0.71 \, \left( {L_{b,K} \over 10^{11} \Lsun}\right)^{0.89} \,
{\rm kpc},
\end{equation}
which we use to set the scale length of the bulges  in our models. Our
results do not depend on this simple  scaling assumption, in practice. 
The main parameter  for the bulge is  its  total mass;  changes in its
actual density distribution are only a second-order effect, and do not
influence our results.

\subsection{Star formation}
\label{sec:sf}

Since the  $K$-band  mass-to-light ratio depends   only  weakly on the
metallicity and age of a stellar population (see \S~\ref{sec:stelpop}), 
and  since  integrated star formation histories of  disk  galaxies are
not expected to exhibit strong variety, the main parameter of interest
is  the  fraction  of  the gas that is  converted into  stars over the
lifetime of the galaxy.

In a   seminal paper Kennicutt  (1989) showed  that  star formation is
abruptly  suppressed below a   critical   surface density,  which   is
associated with the  onset of large-scale gravitational instabilities,
and which can be modeled by a simple Toomre disc stability criterion:
\begin{equation}
\label{Toomre}
\Sigma_{\rm crit}(R) = {\sigma_{\rm gas} \, \kappa(R) 
\over 3.36  \, G \, Q}
\end{equation}
(Toomre 1964).   Here $Q$  is  a   dimensionless constant  near unity,
$\sigma_{\rm gas}$ is the velocity dispersion of the gas, and $\kappa$
is the epicycle frequency given by
\begin{equation}
\label{epicycle}
\kappa(R) = \sqrt{2} \; {V_c(R) \over R} \left( 1 + {R \over V_c(R)} 
{{\rm d}V_c    \over {\rm d}R} \right)^{1/2}.
\end{equation}
This critical  density sets the fraction  of the gas  that is eligible
for star  formation  (see  also Quirk  1972).   Given  the exponential
surface density of  the gas disk, with  scale length $R_d$ and central
surface density $\Sigma_0$,  the radius  $R_{\rm  SF}$, at which   the
density of the disk equals $\Sigma_{\rm crit}$ can be calculated.  The
disk  mass inside   this   radius, and   with surface    density above
$\Sigma_{\rm crit}$ is considered eligible   for star formation   (cf.
Quirk 1972; Kauffmann  1996).   Unless   stated otherwise  we    adopt
$\sigma_{\rm gas} =  6 \kms$ and $Q =  1.5$, as these values have been
shown by   Kennicutt (1989)  to  yield values   of  $R_{\rm SF}$  that
correspond to the radii at which star formation is truncated.

What  remains is   to  quantify the  fraction,  $f_{\rm  SF}$, of this
eligible mass that  is actually turned into stars  over the age of the
galaxy. The star formation  rate of disk galaxies  is well fitted by a
simple Schmidt (1959) law:
\begin{equation}
\label{Schmidt_law}
\psi_{\rm SFR} = a \, \Sigma^{n}_{\rm gas} 
\end{equation}
For a large sample of disk galaxies, Kennicutt (1998)  found $a = 0.25
\pm 0.07 \Msun {\rm pc}^{-2} {\rm Gyr}^{-1}$ and $n=1.4 \pm 0.15$.  If
we  assume that a fraction $\epsilon_{\rm  SF}$  of the mass formed in
new stars is (instantaneously) returned  to the gas  phase by means of
stellar winds and  supernovae, one obtains for  the gas density at the
present
\begin{equation}
\label{SB_gas}
\Sigma_{\rm   gas}(R) =  \left[-{A_{\rm  SF}\over   m} t_{\rm gal}   +
  \left(\Sigma_0 {\rm e}^{-R/R_d} \right)^{1/m} \right]^{m}
\end{equation}
(cf.  Heavens \& Jiminez   1999),   with $m=1/(1-n)$, $A_{\rm SF}    =
(1-\epsilon_{\rm SF}) \, a$, and $t_{\rm gal}$ the  age of the galaxy,
which we define as   the time since the   halo's collapse.  If we  now
couple  this to the threshold  density criterion, we can calculate the
total mass in stars formed in the disk. If  we further assume that all
the gas that is transformed into a bulge component  forms stars at 100
percent efficiency, we obtain
\begin{equation}
\label{starmass}
M_{*} = M_b + 2 \pi\int\limits_{0}^{R_{\rm SF}} \left[\Sigma_0
{\rm e}^{-R/R_d} - g(R)\right] R \, {\rm d}R
\end{equation}
with
\begin{equation}
\label{g_of_R}
g(R) = \left\{ \begin{array}{ll}
       \Sigma_{\rm gas}(R) & \mbox{$\;\;\;\;\;
       ( \Sigma_{\rm gas} > \Sigma_{\rm crit} )$} \\
       \Sigma_{\rm crit}(R) & \mbox{$\;\;\;\;\; 
       ( \Sigma_{\rm gas} \lte \Sigma_{\rm crit})$}
       \end{array} \right.
\end{equation}
Thus, the surface density of the gas is not  allowed to drop below the
critical density given by equation~(\ref{Toomre}).

The total $K$-band luminosity  of  the disk-bulge system then  follows
from $L_K  =   M_{*}/\Upsilon^{*}_K$, where   $\Upsilon^{*}_K$  is the
mass-to-light ratio of the stellar mass.

\subsection{Feedback from supernovae}
\label{sec:feedback}

The dominant source of feedback  that can keep gas out  of the disk is
provided by supernovae (SN). A mass $M_{*}$  of stars produces a total
amount of energy by SN equal to
\begin{equation}
\label{SN_energy}
E_{\rm total} =  \eta_{\rm  SN} \, M_{*} \, E_{\rm SN}
\end{equation}
Here $E_{\rm  SN}$ is the  energy produced  by  one SN, and $\eta_{\rm
  SN}$ is the  number of SN expected per  solar mass of stars  formed. 
Throughout we adopt  $\eta_{\rm  SN} =  4 \times 10^{-3}  \Msun^{-1}$,
consistent with a Scalo (1986) initial mass function (IMF).

In what follows, we  assume that this amount of  energy can  prevent a
mass $M_{\rm  hot}$ from ending   up in the final  disk-bulge  system,
either by means of  completely expelling the gas from  the halo, or by
keeping gas in the halo at the virial temperature, thus preventing its
collapse  to the disk.  Taking account  of the escape  velocity of the
halo, and requiring energy balance, one obtains
\begin{equation}
\label{mass_hot}
M_{\rm hot} =  {M_{*} \,   \varepsilon_{\rm SN}  \, \eta_{\rm  SN} \,
  E_{\rm SN} \over V_{200}^2}
\end{equation}
(cf. Kauffmann \etal 1993; Natarajan  1999), where we introduce a  new
parameter, $\varepsilon_{\rm   SN}$,    which is  a  measure   of  the
efficiency with which the SN energy can prevent the gas from ending up
in the disk  or  bulge.   One might imagine  that  the  free parameter
$\epsilon_{\rm SN}$ is not constant for all galaxies, and we therefore
introduce the following simple scaling relation
\begin{equation}
\label{SN_eff}
\varepsilon_{\rm SN} =  \varepsilon_{\rm SN}^0 \left(V_{200} \over 
250 \kms\right)^{\nu},
\end{equation}
We can now define the fraction
\begin{eqnarray}
\label{Afeedback}
\lefteqn{{M_{\rm hot} \over M_{*}} = 
3.22 \, \varepsilon_{\rm SN}^0 \times} \;\;\; \nonumber \\
& \left({\eta_{\rm SN} \over 0.004 \Msun^{-1}}\right)
\left({E_{\rm SN} \over 10^{51} {\rm erg}}\right) 
\left({V_{200} \over 250 \kms}\right)^{\nu - 2}.
\end{eqnarray}
Thus, in a halo with $V_{200} = 250 \kms$, a mass $M_{*}$ in stars can
prevent a mass $3.22 \, \varepsilon_{\rm SN}^0 \, M_{*}$ from settling
in the disk or bulge.  Stellar winds from  evolving young stars add to
the energy released      by SN, and   we  therefore   allow values  of
$\varepsilon_{\rm SN}$ in excess of unity.
 
We  can now   combine  the    star  formation  recipe  outlined     in
\S~\ref{sec:sf} with  this feedback model to  determine the  masses in
the different  phases:  $M_{*}$, $M_{\rm  hot}$,  and  $M_{\rm cold}$,
where the latter is the amount  of (cold) gas in  the disk that is not
turned into stars. Mass conservation requires  $M_{*} + M_{\rm cold} +
M_{\rm hot} = M_{\rm baryons}$.  Defining  $\epsilon_{\rm gf}$ as  the
galaxy  formation efficiency, which  describes   what fraction of  the
baryonic mass  inside the halo  ultimately ends up in  the disk/bulge,
i.e.,
\begin{equation}
\label{mgal}
M_d + M_b = \epsilon_{\rm gf} \, f_{\rm bar} \, M_{200},
\end{equation}
we can write
\begin{equation}
\label{eps_gf}
\epsilon_{\rm gf} = \left[ 1 + {M_{\rm hot} \over M_d + M_b}
 \right]^{-1}.
\end{equation}
This equation can not be solved analytically,  as $M_{\rm hot}$, $M_d$
and  $M_b$ all depend on $\epsilon_{\rm  gf}$.  However, as we outline
in \S~\ref{sec:outline}, we can solve for $\epsilon_{\rm gf}$ using an
iterative procedure.

Since circular velocities  depend  on whether  or  not the hot  gas is
expelled from the halo,   we  introduce the parameter $f_{\rm   esc}$,
which describes  what fraction of  $M_{\rm hot}$ actually  escapes the
halo.  The remaining fraction  is assumed to  be at virial temperature
and to follow  the same  density  profile as the  dark  matter.  It is
straightforward to show that
\begin{equation}
\label{vcratio}
{V_c(f_{\rm esc} = 1) \over V_c(f_{\rm esc} = 0)} < \sqrt{1 - f_{\rm
    bar}}.
\end{equation}
This  upper  limit is reached  in the  limit  where $\epsilon_{\rm gf}
\rightarrow 0$, and thus  applies to very  low mass systems for  which
the SN feedback  is most efficient.  For  a baryon fraction of $f_{\rm
  bar} = 0.05$,  neglecting $f_{\rm esc}$   thus yields errors  on the
circular velocities that never exceed 3 percent.  Lacking a physically
well-motivated value for  $f_{\rm  esc}$ and given its  insignificance
for the results  presented in this  paper, we set  $f_{\rm esc} = 0.5$
throughout, i.e., midway its two extremes.

\subsection{Stellar populations}
\label{sec:stelpop}

In order to convert stellar masses to luminosities  we need to adopt a
mass-to-light ratio for  the stellar population, $\Upsilon^{*}_K$.  In
the near-infrared the mass-to-light ratios  of stellar populations are
fairly independent of  metallicity   and age, but  these  dependencies
increase strongly   going  towards  bluer passbands  (e.g.,   Maraston
1998a,b).   By focusing on the near-infrared   TF relation we are thus
less sensitive  to  uncertainties related  to  stellar populations (or
extinction  by dust).  Using   the latest version   of the Bruzual  \&
Charlot (1993) population   synthesis models, we  compute the $K$-band
mass-to-light ratio of a stellar population with  a Scalo IMF that has
been producing stars  with a constant rate  over the past  $10$ Gyr (a
typical value for the age of galaxies since collapse). We find a value
of $\Upsilon^{*}_K = 0.4$ and, unless  stated otherwise, we adopt this
value throughout.

\subsection{Constructing a catalogue of model galaxies}
\label{sec:outline}

We now combine all the recipes outlined above to construct a catalogue
of     model    galaxies.    After    choosing   a    cosmology   (see
Table~\ref{tab:cosmo}) and setting the  free parameters in  our model,
we proceed as follows:

First we randomly assign the halo a mass in the range $6.3 \times 10^9
\, h^{-1} \, \Msun \leq M_{200}  \leq 3.6 \times  10^{12} \, h^{-1} \,
\Msun$, corresponding to $30 \,  \kms  \leq V_{200} \leq 250 \,
\kms$. We use the procedure outlined  in the Appendix of Navarro
\etal (1997) to compute  the halo's collapse redshift, $z_{\rm
coll}$, from which we calculate  the age $t_{\rm   gal}$ and the  halo
concentration  parameter $c$.  Next  we randomly draw  a value for the
halo's  spin    parameter   from  the  probability   distribution   of
equation~(\ref{spindistr}).

We  start by assuming  that all  the baryons settle  in  a disk (i.e.,
$\epsilon_{\rm gf} = 1$ and  $M_b = 0$).  Taking adiabatic contraction
into account, we    compute   the scale-length  and central    surface
brightness of the disk (see Mo \etal 1998, and van  den Bosch 1998 for
details). Next we use  equation~(\ref{stabalpha}) to check whether the
disk   is  stable.  If  $\alpha    > \alpha_{\rm crit}$,  an iterative
procedure is  used to compute  the bulge  mass  that yields $\alpha  =
\alpha_{\rm   crit}$  (each iteration    step,  the procedure for  the
adiabatic contraction  is  repeated).   After convergence, we  compute
$M_{*}$ and $M_{\rm hot}$, from which we obtain a new estimate for the
galaxy formation efficiency,   $\epsilon_{\rm gf}$.   We  iterate this
entire procedure  until $M_{*} + M_{\rm cold}  + M_{\rm hot}$ is equal
to   the total baryonic mass,   yielding  a completely self-consistent
disk-bulge-halo model.

Finally  we extract the observational  quantities that enter in the TF
relation.   The luminosity of the  disk  in the $K$-band is determined
from  $M_{*}$ using   the  mass-to-light ratio  $\Upsilon^{*}_K$.   As
velocity  measure, we use the rotation  velocity  at the last measured
point  of the  HI rotation curve  of the  model galaxy.  For  observed
galaxies, this point basically coincides   with the position at  which
the HI column  density, N[HI], falls  below the detectability limit of
the observations, and  is typically   of  the order of $10^{20}   {\rm
  cm}^{-2}$.  Since this column  density is a projected  quantity, the
radius of the last measured point depends on  both the surface density
of  the   disk,   and its   inclination angle  with    respect  to the
line-of-sight.  We therefore draw  a random inclination angle, $i_{\rm
  min} \lte i \lte i_{\rm  max}$, and calculate the circular velocity,
$V_{\rm obs}$, at the radius where  the projected HI column density is
equal to $10^{20} {\rm  cm}^{-2}$.  As mentioned  in \S~\ref{sec:obs},
the     TF  relation    whose    origin    we   seek   to   understand
(equation~[\ref{TFfund}]) is   based on    $V_{\rm flat}$,  i.e.,  the
velocity   at  the   flat      part of   the  rotation     curve.   In
Figure~\ref{fig:xinc} we  show that the  velocity at  the radius where
${\rm N[HI]} =  10^{20} {\rm  cm}^{-2}$ is  a  good representation  of
$V_{\rm flat}$, justifying our choice of velocity measure.

\placefigure{fig:xinc}

Throughout we assume that $\Sigma_{\rm gas} =  1.3 \Sigma_{\rm HI}$ to
take  account of the mass of  helium.  Unless  stated otherwise we set
$i_{\rm  min} =  45^{\rm o}$ and  $i_{\rm  max} = 90^{\rm  o}$ as this
corresponds to   the  typical values  that  observers use   to  select
galaxies  for  their   TF   samples.    Since  the   TF  relation   of
equation~(\ref{TFfund}), which   we use  to constrain our   models, is
based  on spirals of  Hubble type  Sb or later,   we only allow  model
galaxies in our sample with bulge-to-disk ratios $B/D \lte 0.2$.

\section{Cosmological constraints from the Tully-Fisher zero-point}
\label{sec:cosmo_cons}

It  is straigthforward to  show   that simple dynamics  predict a   TF
relation with a slope of $-7.5$ (Dalcanton \etal  1997; White 1997; Mo
\etal 1998; van den Bosch 1998; Syer, Mao \& Mo  1999). For the baryon
density we adopt here  ($\Omega_{\rm  bar} =  0.0125 \,  h^{-2}$)  the
predicted TF can be written as
\begin{equation}
\label{TFpred}
L_K = 7.73 \times 10^{10} {\epsilon_{\rm gf} \over h^3 \, \Omega_0}  
  \left({\Upsilon_K \over 0.3}\right)^{-1} 
  \left({V_{200} \over 200 \kms}\right)^{3},
\end{equation}
(cf. equation~[28] in  van den Bosch 1998).   The observed $K$-band TF
relation of equation~(\ref{TFfund}) in power-law form reads
\begin{equation}
\label{TFobs_lin}
L_K = 5.68 \times 10^{10} \, h^{-2} \left({V_{\rm flat} \over  
  200 \kms}\right)^{4.2}.
\end{equation}
Upon   equating  the  zero-points of the   observed  and  predicted  TF
relations at $V_{200} = 200 \kms$ we obtain
\begin{equation}
\label{combined}
\Omega_0 \, h = 1.363 \, \epsilon_{\rm gf} \, 
  \left({\Upsilon_K \over 0.3}\right)^{-1} \,
  \left({V_{\rm flat} \over V_{200}}\right)^{-4.2}. 
\end{equation}
This equation immediately  suggests that by setting  reasonable limits
on $\epsilon_{\rm  gf}$, $\Upsilon_K$, and $V_{\rm  flat}/V_{200}$, we
can  place strong upper bounds  on $\Omega_0 \, h$.  The mass-to-light
ratio $\Upsilon_K$ in equations~(\ref{TFpred}) and~(\ref{combined}) is
defined as the  ratio of mass  over $K$-band luminosity  for the disk. 
Since not all the mass of the disk is  transformed into stars, this is
not equal to the  {\it stellar} mass-to-light ratio  $\Upsilon_K^{*}$,
i.e.,
\begin{equation}
\label{mtols}
\Upsilon_K = \Upsilon_K^{*} \, {M_d \over M_{*}} 
\end{equation}
and in general  $\Upsilon_K > \Upsilon_K^{*}$.   Using the  Bruzual \&
Charlot (1993) stellar population models, we find that $\Upsilon_K^{*}
\gta 0.3 \Msun/\Lsun$ for a Scalo IMF, at  least if the $K$-band light
is not dominated by  a young ($\lta  1$ Gyr) population.  The $K$-band
mass-to-light ratio depends rather strongly on IMF  (much more so than
on age  and/or   metallicity).   However, most  other   IMFs  that are
frequently used, such as the Salpeter IMF,  have higher mass fractions
at the low mass end, and  therefore yield higher mass-to-light ratios. 
We   can thus consider $\Upsilon_K   = 0.3 \Msun/\Lsun$ a conservative
lower   limit.     Together  with   the    physical  constraint   that
$\epsilon_{\rm gf} \leq 1$, we obtain\footnote{Since the predicted and
  empirical TF relations have different slopes  of $-7.5$ and $-10.5$,
  respectively, the numerical value  of the constraint depends  on the
  velocity  at   which we  compare  the  zero-points.  For  $V_{200} =
  250\kms$ the value of 1.363 reduces to  1.044.  Therefore, by making
  the comparison at $V_{200} = 200\kms$ we are being conservative}
\begin{equation}
\label{constraint}
\Omega_0   \,  h   \leq     1.363   \,  \left({V_{\rm flat}      \over
    V_{200}}\right)^{-4.2}.
\end{equation}
The ratio $V_{\rm flat} / V_{200}$ depends on the concentration of the
dark halo, $c$, which,  for  given $\Omega_0$, $\Omega_{\Lambda}$  and
$h$, depends on  the normalization of the  power spectrum, $\sigma_8$. 
Higher values of  $\sigma_8$ yield more  centrally concentrated halos,
giving     rise  to higher  values  of    $V_{\rm  flat}  /  V_{200}$. 
Equation~(\ref{constraint}) thus yields  an empirical constraint  on a
combination of $\Omega_0$, $\Omega_{\Lambda}$, $h$, and $\sigma_8$.

The constraint of equation~(\ref{constraint}) is over-conservative, in
that {\it all}  baryons  available in  the dark halo  are  turned into
stars.   However, this leaves no HI  gas in  the galaxies, contrary to
observations.  Using  the models  described in  \S~\ref{sec:model}, we
may derive  a  more realistic   prediction  for the  TF relation    of
equation~(\ref{TFpred}).   In   order  to be  conservative  we  ignore
feedback (i.e., we set $\epsilon^0_{\rm SN} =  0$), we assume that all
the gas above the threshold density for star  formation is turned into
stars with $100$ percent  efficiency (i.e., $A_{\rm  SF}$ is such that
$f_{\rm SF}  = 1.0$), and we  adopt the minimal mass-to-light ratio of
$\Upsilon^{*}_K   =  0.3 \Msun /  \Lsun$.   We   thus  aim at creating
maximally bright galaxies.

\placefigure{fig:TFSCDM}

In  Figure~\ref{fig:TFSCDM} we plot  the  $K$-band TF  relation for  a
sample   of  model       galaxies    in a    SCDM      cosmology  (see
Table~\ref{tab:cosmo}).  The parameters  of this model, which we refer
to as model S1, are  listed in Table~\ref{tab:param}. The empirical TF
relation is significantly  steeper ($b=-10.5$ as compared to  $b=-8.9$
for model S1). At the bright end, the model  galaxies are too faint by
$\sim 0.61$ mag at $V_{\rm obs} = 200 \kms$.   Since the parameters of
model  S1 were chosen to ensure  maximally bright galaxies, this rules
against the SCDM cosmology.  For the  galaxies in model  S1 we find an
average of $\langle V_{\rm obs}/V_{200} \rangle  = 1.49 \pm 0.11$, and
from equation~(\ref{constraint})  it is  immediately apparent that the
SCDM model is therefore ruled out.

\placefigure{fig:cons}

The average concentration  of  dark halos  (and therewith the  average
value of $V_{\rm obs}/V_{200}$) decreases  with decreasing $\sigma_8$. 
The TF zero-point thus yields  an upper limit  on $\sigma_8$.  Results
for  a variety  of  ($\Omega_0$,  $\Omega_{\Lambda}$, $h$)-models  are
presented  in  Figure~\ref{fig:cons}.    In   a flat   universe   with
$\Omega_{\Lambda} = 0$, the zero-point of the  TF relation puts severe
limits on the  value of  $\sigma_8$:  for $h=0.5$ we  obtain $\sigma_8
\lta 0.08$.  Such small values are clearly inconsistent with the other
constraints on $\sigma_8$, and we thus conclude that the TF zero-point
can be added to a  long list of  observations that rule against a SCDM
universe.   Flat universes   with $\Omega_0 \lta   0.3$  are far  more
successful. This owes in  large to the  fact that the universal baryon
fraction increases with decreasing $\Omega_0$.

The values of  $V_{\rm obs} /  V_{200}$ found in  our models depend on
the density profiles of the dark halos, and  are thus model dependent. 
Although the  exact density  profiles in  the central regions  of dark
matter halos are still under debate (e.g.,  Moore \etal 1998; Kravtsov
\etal 1998),  most studies now  seem to confirm  that  the {\it outer}
density profiles  fall off as $r^{-3}$. A  generic  property of such a
density profile is  that $V_{\rm obs}  / V_{200}$ will  be larger than
unity, and we  therefore  conclude that  the model  dependency  of our
results is only  weak.  Furthermore, if   halo densities have  central
cusps that are steeper than for the  NFW profiles (as suggested by the
high resolution  simulations of Moore  \etal  1998), our estimates  of
$V_{\rm obs}   / V_{200}$ are too   low, and our limits  on $\sigma_8$
conservative.

\subsection{Comparison with previous work}
\label{sec:cosmo_comp}

The result that the SCDM cosmology  yields galaxies that are too faint
for their rotation velocities has already been pointed out by numerous
previous studies.  However, the analysis presented  here is  very much
different, and yields important new results.

Already the   very first  applications  of   semi-analytical  modeling
revealed problems with  the  normalization of  the TF  relation  (Cole
1991;   White \& Frenk   1991; Lacey \& Silk  1991;  Lacey \etal 1993;
Kauffmann \etal 1993; Cole \etal 1994; Heyl  \etal 1995).  If the star
formation  and feedback parameters of  these models were  tuned to fit
the luminosity function of galaxies, the zero-point of the TF relation
turned  out too  faint by   $\sim 2$  orders  of  magnitude.  This was
interpreted   as indicating that there  are  too many dark  halos of a
given velocity.

Here    we have shown  that   if  one adheres  to the  nucleosynthesis
constraints  on the baryon density,  it is  in principle impossible to
fit the TF zero-point for a SCDM cosmology with a reasonable value for
$\sigma_8$: even after setting the model parameters to yield maximally
bright galaxies, they turn out too faint,  {\it independent of whether
  or  not  these models   fit the galaxy   luminosity  function}.  The
constraints   presented  here are  thus more   strict  than those from
previous SAM studies.

Contrary to all previous SAMs,  Somerville \& Primack (1998) were able
to simultaneously reproduce the  normalization of the TF  relation and
the galaxy  luminosity function in a  SCDM cosmology (with $\sigma_8 =
0.67$). The reason for this apparent contradiction with our results is
two-fold.  First of  all, Somerville \&  Primack used a baryon density
that is $1.5$ times the value adopted here.  Secondly, they considered
isothermal halos,   ignored adiabatic   contraction,   and  made    no
distinction  between $V_{200}$ and  the  actual velocity  measure that
enters  the empirical  TF  relation (the same   is true  for all other
previous  SAMs).   We  have shown   that  a  proper  treatment of  the
different velocities is  a crucial aspect  of  any attempt to try  and
comprehend  (the   origin of) the TF   relation.   In particular, from
equation~(\ref{constraint}) it is immediately   apparent that had   we
directly associated $V_{200}$ with  the rotation measure $V_{\rm obs}$
we would not have been able to rule against SCDM.

In    addition  to  the   SAMs,  several   studies based  on numerical
simulations  have encountered similar  problems with the normalization
of  the TF relation,  e.g.,  Steinmetz \& Navarro  (1999) and Elizondo
\etal   (1999). These   authors  suggest that    the  angular momentum
catastrophe and   declining star  formation  rates,  respectively, are
likely to  be the cause for  this  discrepancy. The analysis presented
here, however, shows that the problem is more fundamental than that.

\section{Constraints from the Slope of the Tully-Fisher Relation} 
\label{sec:slope}

Simple dynamical arguments predict  a   TF relation  with a slope   of
$-7.5$, very different    from the observed  value  of  $-10.5$.  From
equation~(\ref{TFpred}) it is  immediately clear that one can ``tilt''
the TF relation to the observed slope by satisfying the condition
\begin{equation}
\label{condition}
{\epsilon_{\rm gf} \over  \Upsilon_K} \,  \left({V_{200}  \over
    V_{\rm obs}}\right)^3 \propto (V_{200})^{1.2}
\end{equation}
Each of the three    parameters on the  left   side of  this  equation
represents  a different  class    of  physics: the  galaxy   formation
efficiency,  $\epsilon_{\rm gf}$, is  determined  mainly  by processes
related to  feedback, the mass-to-light  ratio $\Upsilon_K$ is related
to  details regarding the  star-formation and stellar populations, and
$V_{200}/V_{\rm obs}$  is  related to the  density  of the dark  halo,
which in  turn reflects  cosmological   initial conditions.   Here  we
investigate which of these processes is  most important in tilting the
TF relation towards its observed slope. Based on the discussion in the
previous \S  , we limit  ourselves here  to our fiducial $\Lambda$CDM3
cosmology (see Table~\ref{tab:cosmo}).

\subsection{Star formation}
\label{sec:sf_slope}

The mass-to-light ratio  $\Upsilon_K$ in equation~(\ref{condition}) is
given by equation~(\ref{mtols}).  The  ratio $M_d/M_{*}$ is set by the
parameters $Q$, which sets  the threshold density for  star formation,
and by $n$     and $A_{\rm SF}$, which  control    the star  formation
efficiency.  In Figure~\ref{fig:slope}  we  plot the TF relations  for
three  models that     only  differ  in   the  value    of    $Q$ (see
Table~\ref{tab:param}).  All three models have  $f_{\rm SF} = 1.0$ and
ignore feedback (i.e., $\epsilon_{\rm  gf} = 1$).   If we set $Q=15.0$
(i.e., ten times its  fiducial value), $\Sigma_{\rm crit}$  becomes so
low that virtually   all the disk material  is  converted into  stars,
independent of $V_{200}$.    Since $V_{200}/V_{\rm obs}$ depends  only
very weakly on $V_{200}$ and $\epsilon_{\rm gf} = 1$  we thus obtain a
TF relation with   a slope of $-7.5$.  We  do not consider a  value of
$Q=15.0$  physical, but merely  show  the results   of this model  for
comparison. Setting $Q$ to  its fiducial value of  $1.5$, $\Upsilon_K$
is   found  to decrease  with increasing   mass.   This  results  in a
significantly steeper TF relation  with $b=-9.1$, and suggests that  a
further decrease of  $Q$ might  actually yield a  TF  relation with $b
\simeq -10.5$.  Indeed,  for $Q=-0.8$ we obtain  $b=-10.0$, consistent
with the observations within  the error-bars.  Besides influencing the
slope of the TF relation, $Q$ also affects the amount of scatter.  For
values of $Q \lta 0.8$ we can obtain even steeper TF relations, but at
the cost of introducing unacceptable levels of scatter.  The origin of
this scatter is discussed in detail in \S~\ref{sec:scatter}.

\placefigure{fig:slope}

We also examined the     influence  of changing the   star   formation
parameters $A_{\rm SF}$ and  $n$ (such as to  obtain $f_{\rm SF} < 1$)
while setting $Q$ to its  fiducial value of $1.5$.   In each case,  we
find    that the Schmidt   law  causes $f_{\rm   SF}$ to decrease with
increasing mass, resulting in a TF relation that is shallower than for
model L1.  In addition, reducing $f_{\rm  SF}$  tends to  increase the
scatter.

\subsection{Stellar populations}
\label{sec:sp_slope}

Next we experiment with a mass dependence of the stellar mass-to-light
ratio    $\Upsilon^{*}_K$.   Rather than adopting  $\Upsilon^{*}_K=0.4
\Msun/\Lsun$, we  set    $\Upsilon^{*}_K  \propto  (V_{200})^s$    and
determine the value  of $s$ that yields  $b=-10.5$,  while setting all
other parameters to that  of model L1. The best  fitting model (L3) is
plotted in the left panel of Figure~\ref{fig:fb} and has
\begin{equation}
\label{ups_mass}
\Upsilon^{*}_K = 0.35 \, \left({V_{200} \over 250 \kms}\right)^{-0.7}.
\end{equation}
This implies $K$-band  mass-to-light ratios for stellar populations in
systems with $V_{200} = 50 \kms$ to be  a factor three larger than for
systems with $V_{200} =  250 \kms$.  Stellar population models suggest
that such  a   large spread  in  $\Upsilon_K^{*}$  cannot   be  due to
variations  in ages   and/or    metallicities, but it might    reflect
systematic variations of the IMF with galaxy  mass (see e.g., Maraston
1998a).  However, there  is   no  observational  support for  such   a
variation.  Furthermore, if low mass  systems have considerably higher
$K$-band  mass-to-light  ratios than  the more   massive systems, this
effect  would  be much stronger  in  bluer  passbands.  One would thus
predict a TF relation  in the $B$-band   with $b \ll -10.5$, in  clear
contradiction with  observations.  We  therefore consider a systematic
variation of $\Upsilon^{*}_K$   with mass an unlikely  explanation for
the steepness of the observed $K$-band TF relation.

\subsection{Cosmological initial conditions}
\label{sec:cosm_condition}

The ratio  $V_{200}/V_{\rm obs}$ in equation~(\ref{condition}) depends
on the concentration $c$ of the dark halo, which is  set by the halo's
collapse redshift $z_{\rm coll}$.  For a  CDM power spectrum, the mass
dependencies of $z_{\rm  coll}$, and henceforth  $c$, are  found to be
fairly small  (e.g.,  Navarro  \etal 1997).    Here  we decouple   the
calculation of  $z_{\rm coll}$ from the  CDM power  spectrum and adopt
the simple parameterization
\begin{equation}
\label{zcoll_param}
1   +    z_{\rm coll}    =   3.5  \,  \left({M_{200}    \over  10^{12}
    \Msun}\right)^{-\zeta}
\end{equation}
with $\zeta$  a free parameter.   The constant $3.5$  is chosen to fit
the TF zero-point for halos with $M_{200} = 10^{12} \Msun$.

We  find that for  $\zeta \simeq   0.26$ the slope    of the model  TF
relation becomes as steep as observed.   This model (L4) is plotted in
the middle  panel of  Figure~\ref{fig:fb}.   From a  comparison of the
mass dependence of $c$ for this  model with that  of a number of scale
free  power spectra in Navarro  \etal (1997), we   infer that $\zeta =
0.26$ corresponds roughly to a  power spectrum of density fluctuations
with an  effective  slope at  the scale of   galaxies of $n_{\rm  eff}
\simeq -1.0$.  This is significantly shallower than for CDM (for which
$n_{\rm eff}$  lies between $-2.0$ and  $-2.4$  for systems with $10^9
\Msun \lte  M_{200} \lte  10^{12}   \Msun$), and therefore  implies  a
cosmology with more  power on small scales  as compared to CDM.  This,
however, seems  inconsistent  with observations  which imply that,  if
anything, the power on small scales has to be {\it  less} than for CDM
(see  e.g., Kauffmann \etal 1993;  Cole \etal 1994; Klypin \etal 1999;
Moore \etal 1999).

We therefore conclude that the slope  of the observed TF relation does
not reflect cosmological initial  conditions, in clear contrast to the
results of AFH98.  This inconsistency is a reflection of the different
choices for the empirical TF relation used to constrain the models: if
AFH98   had  chosen to   use  the  TF relation    of V97  as empirical
constraint, they also would  have concluded that the  slope of the  TF
relation   does  not  reflect cosmological   initial  conditions.  The
empirical relations used by  AFH98 have slopes of  $b \sim -8$ and are
based on HI   line widths and  $H$-band  aperture  photometry.  Subtle
effects can  cause non-linearities  between aperture luminosities  and
total luminosities, which is not taken into account in the modeling of
AFH98. In addition, HI linewidths are difficult  to interpret in terms
of fundamental  parameters  of the galaxy  (see \S~\ref{sec:obs}). The
$K$-band TF  relation used to constrain  our models  is based on total
luminosities derived from CCD   imaging,   and on  rotation   measures
derived from the  complete HI rotation   curves.  We extract the  same
measures from  our model galaxies  ensuring a  fair comparison between
models and observations.  We therefore feel confident that we can rule
out  initial  cosmological conditions  as  the  primary cause  for the
observed slope of the TF relation (at least for a CDM power spectrum).

\subsection{Supernova feedback}
\label{sec:fb_slope}

Having  ruled  out stellar    populations  and  cosmological   initial
conditions, we finally resort  to SN feedback  to try and obtain a  TF
relation as steep as observed.

\placefigure{fig:fb}

The right panel of Figure~\ref{fig:fb}  plots the TF relation of model
L5, for  which we have  tuned the  feedback parameters by  fitting the
empirical  $K$-band TF  relation   (see Table~\ref{tab:param} for  the
parameters). The model fits the empirical TF relation of V97 extremely
well.  We experimented with a variety  of models with different values
of  $\varepsilon_{\rm    SN}^0$  and   $\nu$  and  found    that other
combinations  yield virtually equally good  fits, whereby lower values
of $\varepsilon_{\rm SN}^0$  require a steeper mass  dependence (i.e.,
smaller  $\nu$).  In general,   we  find  that for  the  $\Lambda$CDM3
cosmology   and $\Upsilon^{*}_K =  0.4 \Msun/\Lsun$  the percentage of
baryons  prevented from ending up  in either the  disk or bulge has to
increase from $\sim 10$ percent for systems like  the Milky Way (i.e.,
$V_{\rm obs} =  220 \kms$) to $\sim  70$ percent for low  mass systems
with $V_{\rm obs} = 50 \kms$.  The particular feedback model used here
induces a curvature in  the TF relation (cf.  Natarajan 1999), but for
$\nu \gta -2$ this has a negligible effect over the magnitude range of
$-19   \geq    M_K   -     5    {\rm log}     h   \geq   -25$     (see
\S~\ref{sec:scatter}). Cole \etal (1994), on the other hand, used $\nu
= -3.5$, resulting in a strongly non-linear TF relation.

\subsection{Different cosmological models}
\label{sec:low_omega}

The  discussion  above  is   based  on the  $\Lambda$CDM3   cosmology. 
Although this model is  consistent  with the normalizations from  both
COBE and the cluster abundances, it  predicts halo concentrations with
$c \simeq  9$  on the scale of  galaxies.   Such high values, however,
yield rotation curves that are too steep to be consistent with several
observed rotation curves   (Navarro 1998).  We therefore  now consider
the  $\Lambda$CDM2   cosmology (see Table~\ref{tab:cosmo}),  which was
shown by  Navarro (1998) to be  consistent with all rotation curves in
his sample. Halos  in  this cosmology typically  have  $c \simeq 3$.

\placefigure{fig:altencosmo}

The panels on the left in Figure~\ref{fig:altencosmo} show the results
for a $\Lambda$CDM2 cosmology with the same parameters as for model L1
(see Table~\ref{tab:param}).  Galaxies are typically  $\sim 2$ mag too
bright  for their  rotation velocities.  This   owes to both the  high
baryon fraction in  this cosmology ($f_{\rm  bar} = 0.25$)  and to the
fact that halos are less concentrated  such that $V_{\rm obs}/V_{200}$
is smaller (i.e., we find  $\langle V_{\rm obs}/V_{200} \rangle = 1.26
\pm 0.09$ as  compared to $\langle  V_{\rm obs}/V_{200} \rangle = 1.47
\pm 0.11$ for model  L1). In the middle panel, feedback is added with
its parameters tuned to  fit the TF  relation.  Within our  simplistic
feedback model we require  $\sim 200$ percent of  all the SN energy to
prevent on the order of 80 to 95 percent of  the baryons from settling
in the disk/bulge system.  It seems  unlikely that the combined effect
of SNe  and stellar winds can  be so efficient  (i.e., see  Mac Low \&
Ferrara 1999).    We can lower  the  feedback efficiencies while still
fitting  the TF  relation by  lowering  the  star formation efficiency
$A_{\rm SF}$   (such  that $f_{\rm  SF} <  1$).   In fact,  we  find a
considerable    amount  of freedom   in   the parameters describing SN
feedback and star formation to  obtain good fits  to the empirical  TF
relation.  In general, however, one requires  either a higher feedback
efficiency,  or   a lower star    formation efficiency  than  for  the
$\Lambda$CDM3 cosmology.

\section{Constraints from the Scatter of the Tully-Fisher Relation}
\label{sec:scatter}

The models in the previous sections contain two  sources of scatter in
the   TF  relations:   the    spread   in    halo   spin   parameters,
$\sigma_{\lambda}$, and  the spread  in inclination  angles,  which we
quantify by $\Delta i \equiv i_{\rm max} - i_{\rm min}$. 

\placefigure{fig:chisto}

For a  halo of a given   mass, a non-zero  $\sigma_{\lambda}$ yields a
spread in both the magnitudes  and rotation velocities of disks. Lower
values of $\lambda$ yield more compact disks with a larger fraction of
their mass eligible for  star formation (i.e., galaxies are relatively
bright). In addition, the compactness of  the disk induces, because of
the adiabatic   contraction, a more   strongly  concentrated halo, and
therefore a relatively high  value of $V_{\rm obs}/V_{200}$.  Model L0
(for which $Q=15.0$)  in Figure~\ref{fig:slope} depicts a large amount
of  scatter.    Its origin  is  illustrated   in  the small   inset in
Figure~\ref{fig:slope},   which shows  the  ``TF relation''  for model
galaxies with the  same mass, but different  spin parameters.  Because
of the  high value of $Q$, the  critical density for star formation is
relatively low.  This implies that  the radius $R_{\rm SF}$ where star
formation  is  truncated occurs at relatively  large  radii of $\sim 8
R_d$, where the   growing curve of   the cumulative disk mass  is very
flat.   Consequently,  scatter  in  $\lambda$ translates   mainly into
scatter in  $V_{\rm obs}$ and not in  $M_K$.  The  opposite applies to
model  L2, for which $Q=0.8$.   Here,  $R_{\rm  SF} \sim  3 R_d$,  and
$\sigma_{\lambda}$ translates  mainly into scatter in disk magnitudes.
Most remarkably,  for our  fiducial  value of   $Q$ (i.e., model  L1),
$\sigma_{\lambda}$ translates into scatter  in both $M_K$  and $V_{\rm
obs}$ but such  that galaxies with  different values  of $\lambda$ are
aligned with the actual TF relation.  Consequently, $\sigma_{\lambda}$
contributes   only weakly  to the  scatter  in the  TF relation.   For
$Q=1.5$ we thus  obtain a natural explanation  for the small amount of
scatter observed.

Different values of  $i$ for otherwise  identical galaxies  results in
different radii where ${\rm N[HI]}  = 10^{20} {\rm cm}^{-2}$, and thus
in    different    values   of    $V_{\rm    obs}$.     As  shown   in
Figure~\ref{fig:xinc}, the rotation  measures   used in the  model  TF
relations are fairly insensitive to $i$,  indicating that the rotation
curves are  close to flat. Consequently,  $\Delta i$ only adds a small
fraction of the total scatter.

An additional  source of scatter,  which  has hitherto not been  taken
into  account, stems from the  variation in mass aggregation histories
(MAHs), which induces scatter in the  density profiles of halos of the
same  mass.   For halos that   are well represented  by a  NFW density
profile,  this implies a scatter in   the concentration parameter $c$. 
Thus if we know how scatter in MAHs translates into  scatter in $c$ we
can investigate how this  cosmological scatter adds  to that of the TF
relation.  This  information    is provided by   the   high resolution
numerical simulations  of Jing (1999).   In our models $c$ is computed
from   the collapse redshift  of the  halo, and we  can thus introduce
scatter in $c$  by  instituting scatter  in $z_{\rm  coll}$.  To  that
extent we introduce  a  new  parameter,  $\sigma_z$,  which sets   the
variance in collapse   redshifts  of a  given   mass, and tune  it  to
reproduce the scatter in $c$ in the simulations of Jing.  Although the
amount   of scatter  increases  strongly   with   the amount  of  halo
substructure, the  fragility of disks suggests  that they are embedded
in relatively   smooth halos.   We therefore normalize   $\sigma_z$ by
fitting the distribution of  concentration parameters in halos without
significant amounts  of substructure.   In Figure~\ref{fig:chisto}  we
plot the log-normal    distribution  that  best fits the     numerical
simulations of Jing, together with  a histogram for  halos in model L5
and  with  $\sigma_z =   0.4$.  Except  for  a  small  offset   in the
zero-point,   which is not   important  for  obtaining  an appropriate
measure for the amount of scatter, the agreement  is very good. We can
thus use $\sigma_z = 0.4$ as a representative measure for the standard
deviation  in    MAHs  for halos surrounding      disk galaxies.  This
translates into scatter in the concentration of halos, and thus in the
ratios $V_{\rm obs}/V_{200}$ and the surface densities of the disks.

\placefigure{fig:scatter}

In order to investigate the main source of scatter in the TF relation,
we compare four TF relations that only differ in the values of $\Delta
i$, $\sigma_{\lambda}$  and $\sigma_z$ (see Figure~\ref{fig:scatter}).
The models  have the  same  parameters as model  L5.   The upper right
panel, for which $\Delta i =  \sigma_{\lambda} = \sigma_z = 0$ clearly
reveals the  small  amount of  curvature  eluded  to above, and  which
causes a non  zero scatter of  $\sigma_M =  0.06$ mag around  the best
fitting linear relation.  The main source of this  curvature is the SN
feedback.  Setting $\sigma_{\lambda}$  to its fiducial value  of $0.6$
increases the scatter  to $\sigma_M = 0.13$ mag.   If we further add a
variation in  inclination angles, with   $\Delta i = 45^{\rm o}$,  the
scatter becomes that of model  L5: $\sigma_M =  0.16$ mag.  If finally
we set $\sigma_z$  to   its fiducial  value  of  $0.4$, this   further
increases $\sigma_M$  to $0.20$ mag.   We find that for $\sigma_z \gta
0.4$, the dispersion  in   halo collapse  redshifts becomes the   main
source of scatter in the TF relation.

V97 derived  confidence levels on the  amount  of intrinsic scatter in
the  TF relation of  equation~(\ref{TFfund}), and found $\sigma_M \lte
0.30$ mag at 95 percent confidence level,  with a most likely value of
$0.11$ mag.  This is in excellent agreement with our value $\sigma_M =
0.20$ mag for $\sigma_z = 0.4$.

\section{Scatter from cosmological initial conditions}
\label{sec:elcomp}

Eisenstein \&  Loeb  (1996; hereafter   EL96)  used the  excursion-set
formalism of Lacey \& Cole (1993) to compute MAHs for halos of a given
mass, and calculated predictions  for the amount  of scatter in the TF
relation  resulting from this  variation in MAHs.  They concluded that
the amount of scatter in  the observed TF   relations is smaller  than
what one expects from  the scatter in MAHs.  However, we have shown in
\S~\ref{sec:scatter} that our models predict an amount of scatter that
is in   good agreement  with  observations, even   after we   take the
variance in MAHs, modeled quantitatively  by $\sigma_z$, into account.
Here we investigate the reason for this disagreement.

In  their  analysis,   EL96   assume that  the   mass  gained   in  an
infinitesimal redshift interval is  accreted in a smooth and spherical
fashion, and calculate  the  induced change  in  binding energy, $E$.  
Using the assumption that halos are isothermal, $E \propto V_{200}^2$,
and any variation  in MAHs thus  produces scatter in  the TF relation. 
EL96  assume that  the velocity  measure  used  in the TF  relation is
strictly  determined by $V_{200}$ and  that the  overall luminosity of
the  galaxy is directly related  to the mass of  its  halo. For a SCDM
model with $\sigma_8 = 0.65$ they find $\sigma_M \simeq 0.36$ mag, and
conclude  that this is  too large to be   consistent with the observed
scatter.

\placefigure{fig:el}

We can make a  rough comparison between our method  and that  of EL96,
using the fact that $\sigma_z$ can  be used to parameterize scatter in
MAHs.   The analysis of EL96 is  based on the spherical collapse model
which,  for halos with  the  same mass and  in  an  Einstein-de Sitter
universe, yields $E \propto (1 + z_{\rm  coll})$.  Combining this with
the assumption of isothermal halos we thus obtain $V_{\rm obs} \propto
(1 +  z_{\rm    coll})^{1/2}$.  In our   analysis,  $\sigma_z$ induces
scatter  in halo   concentration   parameters, which  in   turn yields
variance in  $V_{\rm   obs}/V_{200}$.   The calculation  of  $c$  from
$z_{\rm coll}$ is based on the Press-Schechter  formalism, and is well
calibrated against    high   resolution $N$-body     simulations.   In
Figure~\ref{fig:el} we compare the distributions of $V_{\rm obs}$ that
results from scatter in MAHs derived using both methods. The method of
EL96   predicts a standard deviation in    rotation velocities that is
approximately a factor  two larger than for  our method.  If we  halve
the amount of scatter expected  by EL96, we  obtain $\sigma_M = 0.18$,
bringing their results in good agreement  with ours.  Since our method
does not make any  assumptions regarding spherical collapse, uses more
realistic halo density profiles, and has been well calibrated, we feel
confident  that our  estimates of   the  expected scatter  in  the  TF
relation are the more reliable.  We therefore  disagree with EL96, and
conclude  that the observed   amount  of scatter  in  TF  relations is
consistent with expectations from cosmological initial conditions.

\section{Constraints from Gas Mass Fractions}
\label{sec:prop}

Our results regarding the  TF relation would be  of little use  if the
model galaxies were   not representative of  real  galaxies.   Here we
briefly examine the gas mass fractions  of our model galaxies, as they
provide useful, additional constraints.

\placefigure{fig:gasfrac}

McGaugh  \&  de Blok  (1997; hereafter MB97)   have  obtained gas mass
fractions, defined by
\begin{equation}
\label{gasmassfrac}
f_{\rm gas} = {M_{\rm cold} \over M_{\rm cold} + M_{*}},
\end{equation}
for a sample of   108 spiral galaxies. In Figure~\ref{fig:gasfrac}  we
plot $f_{\rm   gas}$ as function  of  absolute magnitude   and central
surface brightness for  the data of  MB97 and for  three of our models
that all fit   the empirical  TF  relation of  V97.  Model L5   agrees
extremely well with the observations, indicating that the observed gas
mass fractions in spirals are consistent with being governed by a star
formation  threshold density  that originates from  Toomre's stability
criterion with $Q=1.5$. Model L2, for which $Q = 0.8$, yields gas mass
fractions that  are slightly too   large. Together with the  fact that
this model predicts a TF zero-point that depends on surface brightness
(see \S~\ref{sec:scatter}) and that $Q=0.8$  is inconsistent with  the
empirically  determined value prompts us to  rule  against this model. 
Model L8, for which $f_{\rm SF} < 1$, predicts gas mass fractions that
are clearly too large to be consistent with observations. We find this
to be  a general problem for  models in which only  a  fraction of the
mass eligible for star formation  is  actually turned into stars  over
the lifetime of the galaxy.  Virtually all models with $Q$ and $A_{\rm
  SF}$  set  to  their  fiducial  values  (that   have been determined
empirically) yield gas mass fractions  in good agreement with the data
of MB97, suggesting that indeed $f_{\rm SF} \simeq 1.0$.

\section{Summary and Conclusions}
\label{sec:concl}

We  have used new semi-analytical models  to investigate the origin of
the near-infrared TF relation.   By focusing  on the near-infrared  we
are  less susceptible to uncertainties  related to  extinction by dust
and stellar populations.  As  empirical  constraint we have  used  the
$K$-band TF relation  of  V97,  which is   the only near-infrared   TF
relation based   on CCD  imaging  (rather than   aperture  photometry)
combined  with full HI rotation curves  (rather than  HI line widths). 
An important improvement  over previous studies aimed at understanding
the TF  relation  concerns our  special  care in  extracting the  same
luminosity and rotation  measures from our model, as  the ones used in
the empirical TF relation.

Since dark  halos have density profiles that  at large radii  fall off
more rapidly than an isothermal (i.e., the NFW profiles used here fall
off as $r^{-3}$), the circular velocities of the disk at the flat part
of     the rotation curve   are  generally   larger than the  circular
velocities at the virial radius.   This has important consequences for
the TF zero-point, as it allows us  to put constraints on cosmological
parameters.  In    a  SCDM universe, dark    halos  are too  centrally
concentrated,  and the baryon  fraction is too  low, to be able to fit
the $K$-band TF relation.  The observed zero-point puts an upper limit
on the normalization of the power spectrum of $\sigma_8 \lta 0.08$, in
clear contradiction with constraints from  COBE and the abundances  of
rich clusters of   galaxies.   We have  shown  that the    observed TF
zero-point (combined   with  the  nucleosynthesis  constraints on  the
baryon density)  favor a Universe  with $\Omega_0  \lta 0.3$. Previous
studies of  the TF   relation based on  SAMs  were  only able to  rule
against SCDM, {\it if the models were normalized to fit the luminosity
function of  galaxies}.  The reason why we  can rule  out SCDM without
this additional constraint is  our more sophisticated treatment of the
relationship between $V_{200}$ and $V_{\rm obs}$.

Whereas simple dynamics predict a TF  relation with a slope of $-7.5$,
the $K$-band TF relation reveals a slope of $-10.5$. This implies that
the physics regulating  star formation and  feedback, coupled with the
mass dependence of halo densities and stellar populations, has to tilt
the TF   relation   to its observed  slope.    The  introduction  of a
stability-related star formation threshold density increases the slope
of the TF relation, reduces its scatter, and yields gas mass fractions
that are  in excellent agreement  with observations.  Setting Toomre's
$Q$   parameter  to its empirically  determined   value  tilts  the TF
relation to a slope   of $\sim -9$, and   additional physics are  thus
required to  obtain  a TF   relation as  steep  as observed.   We have
presented four different physical mechanisms  that all yield good fits
to  the  observed $K$-band TF relation:  lowering  $Q$  to $\sim 0.8$,
systematic variations in   the $K$-band mass-to-light ratios, a  power
spectrum of  initial density fluctuations that  differs from  that for
CDM,  and feedback. Except  for the  latter,  each of these results in
inconsistencies with other  observations, and can  be ruled out as the
prevailing  mechanism   for   the  observed  characteristics    of the
near-infrared TF relation.

The  feedback efficiency required to   tilt   the TF relation to   its
observed slope has  to be  higher in lower  mass  systems, and depends
strongly on $\Omega_0$ and $h$ (if the baryon fraction  is taken to be
constrained by the nucleosynthesis results).  In particular, in a flat
universe with $\Omega_0 = 0.2$ and $h=0.5$  (a cosmological model that
yields low density halos in agreement  with observed rotation curves),
$\sim 200$ percent  of the available SN  energy is required to fit the
observed  TF zero-point.  Although in principle  unphysical, we do not
wish to draw too strong conclusions  from our oversimplified model for
feedback.  We do  emphasize, however, that some  amount of feedback is
required  to  yield TF relations  with a  slope as steep  as observed. 
More sophisticated models of SN feedback,  such as in  the work of Mac
Low \&  Ferrara   (1999), are required    to  investigate whether  our
inferred feedback efficiencies are realistic.  Our best fitting models
with feedback reveal  a small amount of  curvature in the TF relation,
in qualitative agreement   with recent observations of  dwarf galaxies
(Mathews, van Driel \& Gallagher 1998; Stil \&  Israel 1998).  Further
constraints  on  the models  come from  the gas  mass fractions, which
require that virtually  all the gas  mass eligible for  star formation
(i.e., with densities above the  critical density) is transformed into
stars over the lifetime of the galaxy.

For  the fiducial  value of $Q$  the scatter  in  halo spin parameters
affects  both the luminosities   and the observed rotation velocities,
but to such   a degree   that  galaxies are  scattered along   the  TF
relation, rather than  perpendicular to it.   The introduction of star
formation threshold densities thus yield a natural explanation for the
small amount of scatter observed. Taking account of a realistic spread
in  halo concentrations, consistent  with  what  one expects from  the
spread in  mass aggregation  histories, our  model that  best fits the
zero-point and slope of the $K$-band TF relation predicts a scatter of
$\sim 0.2$  mag only, in  excellent agreement  with observations. In a
follow-up  paper (van  den Bosch \&    Dalcanton 1999) we compare  the
models presented here to numerous other observational constraints.


\acknowledgments

This work has   benefited   greatly from discussions  with    Marcella
Longhetti, Claudia Maraston, Jeroen   Stil, and Marc Verheijen.  I  am
indebted to Julianne Dalcanton and George Lake for advice and critical
assessments  of an earlier draft  of the  paper, to St\'ephane Charlot
for providing  results from his stellar  population models, and to the
anonymous   referee  for  his    comments  that  helped   improve  the
presentation of the paper.  Support for this work was provided by NASA
through  Hubble Fellowship grant  \#  HF-01102.11-97.A awarded by  the
Space Telescope Science Institute, which  is operated by AURA for NASA
under contract NAS 5-26555.


\ifemulate\else
  \baselineskip=10pt
\fi


\clearpage


\ifsubmode\else
\baselineskip=14pt
\fi


\newcommand{\figcapxinc}{The ratios $V_{\rm obs}/V_{200}$ (left panel)
  and $R_{\rm  obs}/R_d$ (right panel)  as functions of the  cosine of
  the  inclination angle  for model  galaxies with   the parameters of
  model L5 (see  Table~\ref{tab:param}). Here $R_{\rm obs}$ is defined
  as the radius where the projected surface  density of the HI reaches
  a column density  of $10^{20}  {\rm  cm}^{-2}$, and  $V_{\rm obs}  =
  V_c(R_{\rm obs})$.    For  disks seen  close to   edge-on ($i \simeq
  90^{\rm o}$), $R_{\rm obs}$ becomes as large as  $\sim 10 R_d$.  For
  close  to  face-on  systems, $R_{\rm  obs}/R_d  \simeq   4.5$ with a
  considerable  scatter.    Despite this  relatively  large  change in
  $R_{\rm obs}$, the ratio $V_{\rm obs}/V_{200}$ is almost independent
  of $i$.  This indicates that $V_{\rm obs}$ is an accurate measure of
  the flat part  of the rotation curve,  and we  therefore use $V_{\rm
    obs}$ and   $V_{\rm  flat}$  without distinction   throughout  the
  paper.\label{fig:xinc}}

\newcommand{\figcapTFSCDM}{Model TF relation  for model  S1 (SCDM with
  parameters  set to obtain  maximally bright galaxies).  Open circles
  correspond to the model galaxies,  and the solid line corresponds to
  the observed  $K$-band TF relation of V97.   Only galaxies with $-19
  \gte M_K - 5 {\rm  log} h \gte  -25$ are shown, which corresponds to
  the magnitude interval on which the  empirical TF relation is based. 
  The slope of  the model  TF  relation  is too  shallow ($b=-8.9$  as
  compared to $b_{\rm obs} = -10.5$), and model galaxies are too faint
  at the bright end.  Since galaxies  are maximally bright, this rules
  against the SCDM model (see text).\label{fig:TFSCDM}}

\newcommand{\figcapcons}{Constraints  on $\sigma_8$ and  $h$  for four
  flat cosmologies with different values of $\Omega_0$. The light gray
  areas correspond to the  parts of parameter  space allowed by the TF
  zero-point.  Cosmological  models with  $(\sigma_8,h)$ in  the white
  areas yield  galaxies that are too  faint.  In addition, constraints
  are plotted from the 4 year data from  the COBE DMR experiment (Bunn
  \& White  1997;  solid lines) and   from the  cluster  abundances at
  $z=0$: $\sigma_8  =      (0.52  \pm 0.04)    \,    \Omega_0^{-0.52 +
  0.13\Omega_0}$ (Eke,  Cole  \& Frenk  1996; dark  gray  bars) and $z
  \simeq 0.7$: $\sigma_8 = (0.8 \pm 0.1) \, \Omega_0^{-0.29}$ (Bahcall
  \&   Fan 1998; median   gray   bars). In   a flat universe   without
  cosmological constant (upper right panel)  the TF zero-point imposes
  stringent constraints on $\sigma_8$ and  $h$. In order for  galaxies
  in this  cosmology to be  as bright as  required by the TF relation,
  $\sigma_8  \lta 0.08$ when $h=0.5$.   This is in clear contradiction
  with the other constraints.  Since a  decrease of $\Omega_0$ induces
  an increase of the  baryon fraction (i.e.,  we adopt $f_{\rm bar}  =
  0.0125  \, \Omega_0^{-1} h^{-2}$   throughout) , the  TF constraints
  become  less  restrictive for lower   values  of the matter density.
  Combining          all   constraints    suggests     $\Omega_0  \lta
  0.3$.\label{fig:cons}}

\newcommand{\figcapslope}{The upper panels show TF relations for three
  models in  a flat    $\Lambda$CDM  universe with $\Omega_0   =  0.3$
  (symbols  are the same as   in Figure~\ref{fig:TFSCDM}).  All  these
  models ignore  feedback (i.e., $\epsilon_{\rm  gf} =1$) and have all
  the mass that is eligible for star formation  turned into stars over
  the lifetime of the galaxy (i.e., $f_{\rm SF} = 1$). The models only
  differ in the value of $Q$ as indicated  in the panels. In addition,
  the  panels list the  model ID and the  slope of the best fitting TF
  relation. The small insets plot TF relations for model galaxies that
  only differ   in $\lambda$ (all  galaxies  have  $M_{200}  = 3\times
  10^{11} h^{-1}\Msun$ and  have a value of  $\lambda$ drawn from  the
  log-normal   distribution    of     equation~[\ref{spindistr}]  with
  $\sigma_{\lambda} =  0.6$). These insets show how $\sigma_{\lambda}$
  contributes to the scatter (see discussion in \S~\ref{sec:scatter}).
  The       lower       panels    plot      ${\rm    log}(\Upsilon_K)$
  (equation~[\ref{mtols}])  as  function of ${\rm  log}(V_{\rm obs})$.
  The dotted  lines correspond to $\Upsilon_K   = \Upsilon_K^{*} = 0.4
  \Msun/\Lsun$.  Lower  mass systems   require a significantly  higher
  value of $\Upsilon_K$ than more massive galaxies in order to fit the
  observed slope.   This can be  accomplished by lowering $Q$ to $\sim
  0.8$.\label{fig:slope}}

\newcommand{\figcapfb}{Model TF  relations for     three $\Lambda$CDM3
  models   that  all  fit  the    empirical  $K$-band TF  relation  of
  equation~(\ref{TFfund}).  Each of these models, however, represent a
  different class  of  physics.   In  model L3,   the $K$-band stellar
  mass-to-light ratios are varied  with mass to  obtain $b=-10.5$.  In
  model L4, the same is accomplished, but by adopting a power spectrum
  that is shallower than for CDM on the scale of galaxies. Finally, in
  model  L5   feedback is included to   tilt  the TF  relation  to its
  observed slope.\label{fig:fb}}

\newcommand{\figcapaltencosmo}{Model  TF    relations       for  three
  $\Lambda$CDM2 models (upper panels).  The  models differ only in the
  parameters  describing   star    formation  and     feedback    (see
  Table~\ref{tab:param}). The effects of  these two physical processes
  are indicated in  the panels  in the two  lower rows,  where we plot
  $\epsilon_{\rm  gf}$   (a measure of   the  feedback efficiency) and
  $f_{\rm   SF}$  (a measure  of  the  star  formation efficiency)  as
  functions of ${\rm log}(V_{\rm obs})$.  If no feedback is taken into
  account, and all the mass eligible for star formation is turned into
  stars over the lifetime of the galaxy (panels on the left), galaxies
  are almost two  magnitudes too bright.   Models L7 and  L8, in which
  feedback  is  taken into  account,  both provide   good fits  to the
  observed TF relation.\label{fig:altencosmo}}

\newcommand{\figcapchisto}{Histogram of the normalized distribution of
  $c/c_0$ for model L5 with $\sigma_z = 0.4$.   Here $c_0$ is the halo
  concentration parameter  predicted  by the Press-Schechter formalism
  (see Navarro  \etal 1997 for  details).  The solid  curve represents
  the log-normal distribution of $c/c_0$ for halos without significant
  amounts of substructure as   determined  by Jing (1999) from    high
  resolution  $N$-body simulations   (in   Jing's notation,  we   only
  included halos with ${\rm dvi}_{\rm max} <  0.15$).  Despite a small
  offset in the mean (Jing finds $\langle c \rangle \simeq 0.95 c_0$),
  the two  distributions agree well, and  we conclude therefore that a
  standard deviation  of $0.4$  in halo collapse   redshifts is a good
  representation of  the scatter in halo   concentrations that owes to
  scatter in mass aggregation histories.\label{fig:chisto}}

\newcommand{\figcapscatter}{Illustration of the build-up of scatter in
  the  TF relation  of model  L5.  Contrary  to previous plots  (i.e.,
  Figures~\ref{fig:TFSCDM},       \ref{fig:slope},       \ref{fig:fb},
  and~\ref{fig:altencosmo}) we plot   the TF relation over the  entire
  range of $M_K$ of our model galaxies.  This emphasizes the curvature
  present in the TF  relation, and which owes mainly  to the feedback. 
  In the upper left panel, no intrinsic scatter is present.  The small
  amount of scatter in the  TF relation of  $\sigma_M = 0.06$ mag owes
  to the  curvature. The upper right panel  shows the same TF relation
  but now  with $\sigma_{\lambda} =  0.6$.  The variance in  halo spin
  parameters increases $\sigma_M$  to 0.13 mag.   Setting  $\Delta i =
  45^{\rm o}$  further increases the scatter  to $\sigma_M = 0.16$ mag
  (lower  left panel).  Finally, the addition   of scatter in the halo
  collapse redshifts   ($\sigma_z=0.4$) yields   a  total scatter   of
  $\sigma_M = 0.20$ mag.\label{fig:scatter}}

\newcommand{\figcapel}{The shaded histogram  plots the distribution of
  ``observed'' velocities (normalized to  the mean) for model S1,  but
  with $\Delta i  = \sigma_{\lambda} =  0$  and $\sigma_z = 0.5$.  The
  thick solid line    corresponds   to the same   distribution,    but
  calculated  from  the procedure used  by Eisenstein  \& Loeb (1996). 
  This   latter method  is based    on the  simplified assumptions  of
  spherical accretion and isothermal halos, and predicts roughly twice
  the amount of scatter in the TF relation. \label{fig:el}}

\newcommand{\figcapgasfrac}{The gas mass  fraction $f_{\rm gas}$  (see
  equation~[\ref{gasmassfrac}])   as  function  of  absolute  $B$-band
  magnitude (left panels) and  central disk surface brightness  (right
  panels).  Upper panels plot  the data of McGaugh  \& de Blok (1997),
  while the panels in the lower three rows correspond to models L2, L5
  and L8, as indicated. The data is  based on $B$-band photometry, and
  we convert our models ($K$-band) to the $B$-band using $B-K = 0.64 -
  0.18 (M_K -  5 {\rm log} h_{75})$ with  $h_{75} = H_0/(75 \kmsmpc )$
  (see Appendix A in van den Bosch \& Dalcanton 1999).  The thin solid
  lines  have no  physical meaning,  but are plotted  to facilitate  a
  comparison   between models and  data.    Model  L5 yields gas  mass
  fractions that are in excellent  agreement with the data.  The model
  reproduces the decrease of $f_{\rm  gas}$ with increasing luminosity
  and surface brightness, and the absolute values of $f_{\rm gas}$ are
  in excellent agreement with the data of MB97.  Models  L2 and L8, on
  the other hand, predict gas mass fractions  that are too large to be
  consistent with observations.\label{fig:gasfrac}}


\ifsubmode
\figcaption{\figcapxinc}
\figcaption{\figcapTFSCDM}
\figcaption{\figcapcons}
\figcaption{\figcapslope}
\figcaption{\figcapfb}
\figcaption{\figcapaltencosmo}
\figcaption{\figcapchisto}
\figcaption{\figcapscatter}
\figcaption{\figcapel}
\figcaption{\figcapgasfrac}
\clearpage
\else\printfigtrue\fi

\ifprintfig

\clearpage
\begin{figure}
\epsfxsize=16.0truecm
\centerline{\epsfbox{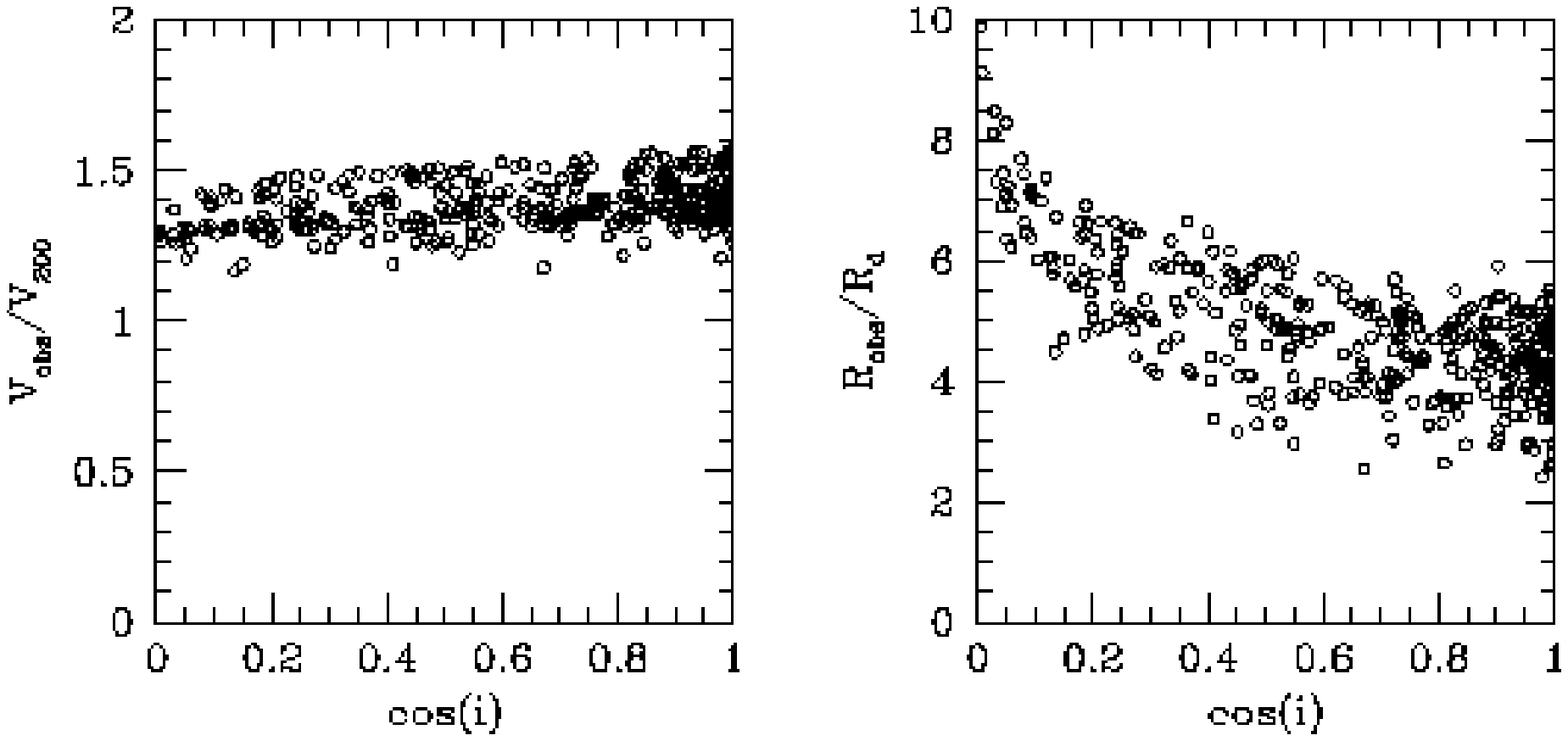}}
\ifsubmode
\vskip3.0truecm
\setcounter{figure}{0}
\addtocounter{figure}{1}
\centerline{Figure~\thefigure}
\else\figcaption{\figcapxinc}\fi
\end{figure}


\ifemulate\else
  \clearpage
\fi
\begin{figure}
\epsfxsize=7.0truecm
\centerline{\epsfbox{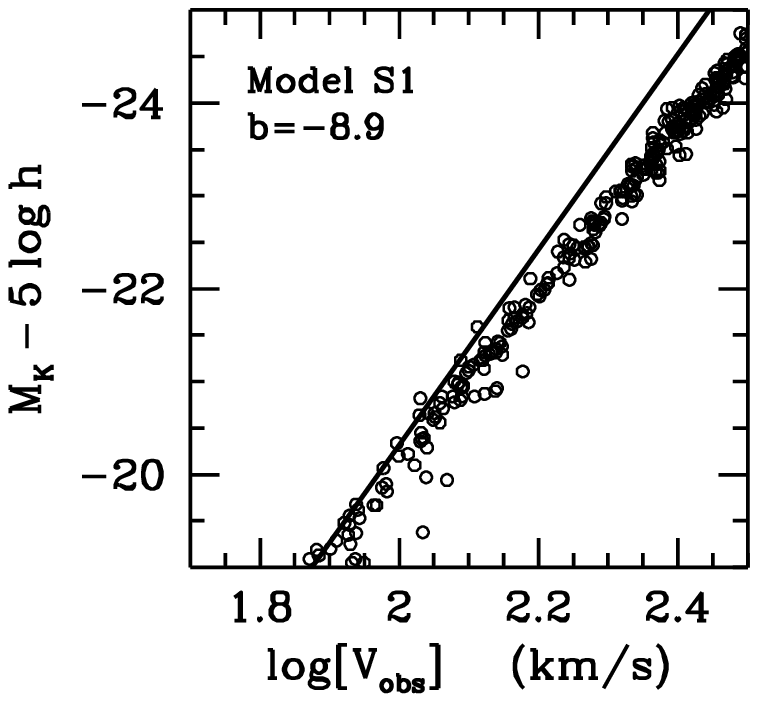}}
\ifsubmode
\vskip3.0truecm
\addtocounter{figure}{1}
\centerline{Figure~\thefigure}
\else\figcaption{\figcapTFSCDM}\fi
\end{figure}


\clearpage
\begin{figure}
\epsfxsize=15.0truecm
\centerline{\epsfbox{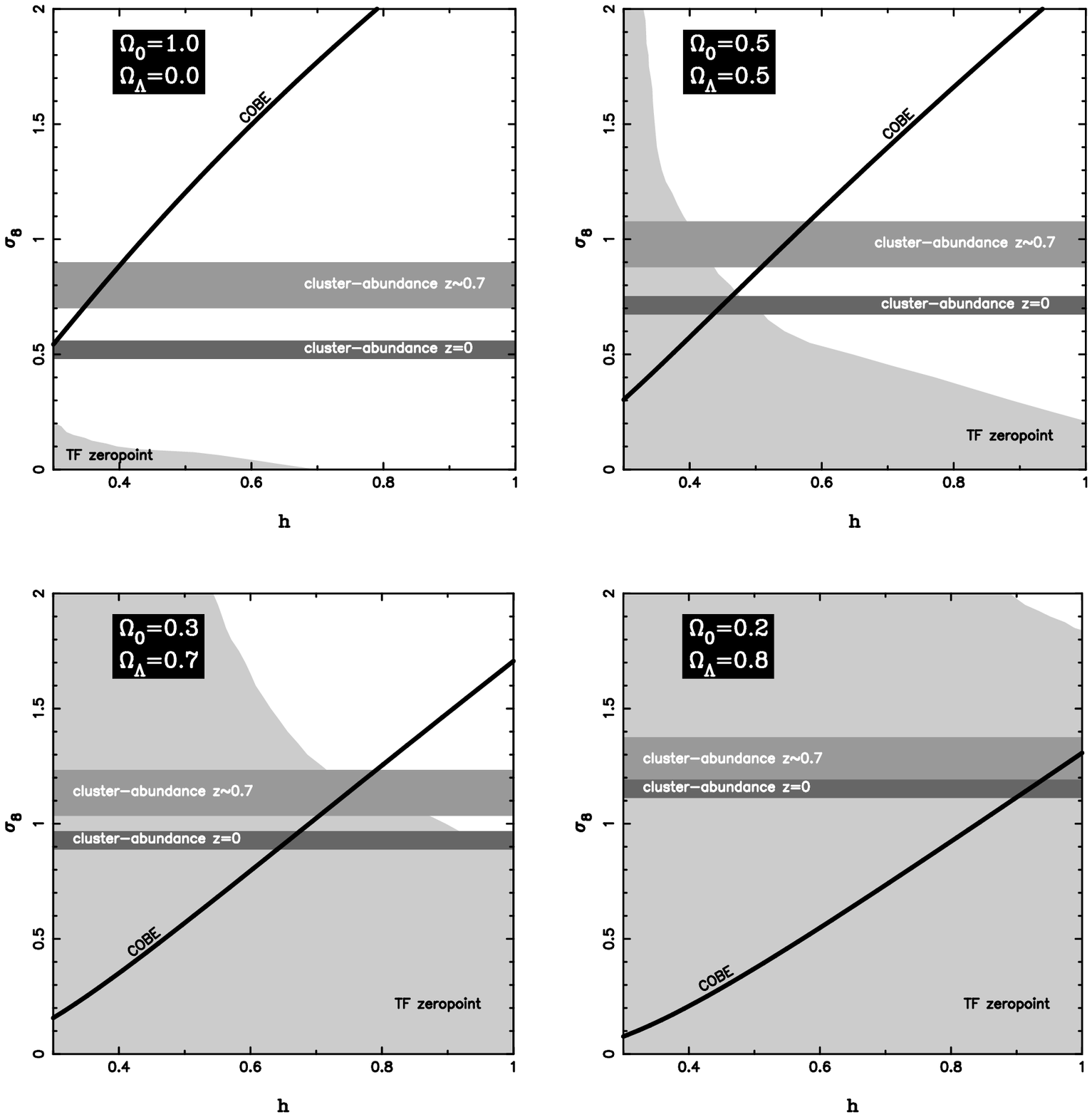}}
\ifsubmode
\vskip3.0truecm
\addtocounter{figure}{1}
\centerline{Figure~\thefigure}
\else\figcaption{\figcapcons}\fi
\end{figure}


\clearpage
\begin{figure}
\epsfxsize=14.0truecm
\centerline{\epsfbox{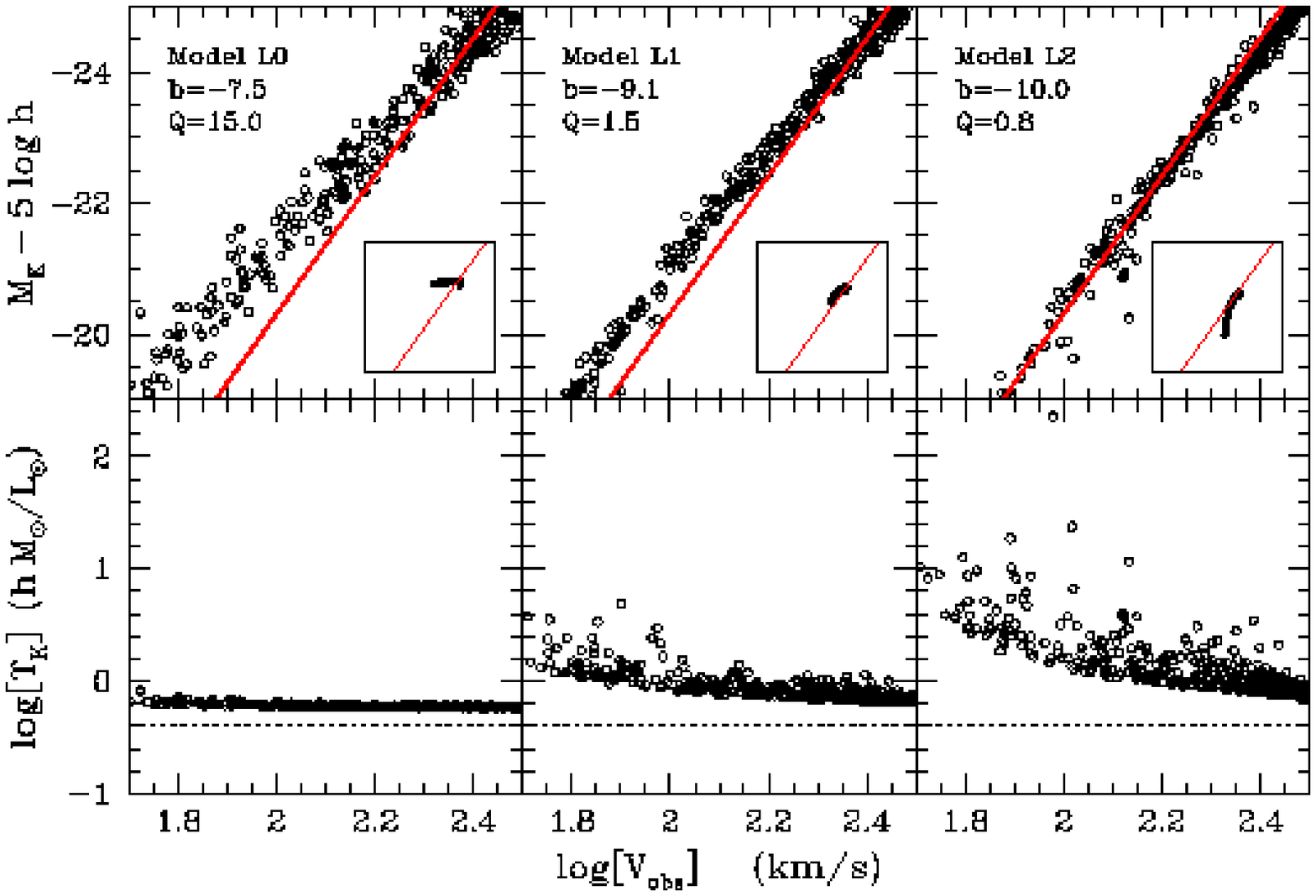}}
\ifsubmode
\vskip3.0truecm
\addtocounter{figure}{1}
\centerline{Figure~\thefigure}
\else\figcaption{\figcapslope}\fi
\end{figure}


\ifemulate\else
  \clearpage
\fi
\begin{figure}
\epsfxsize=16.0truecm
\centerline{\epsfbox{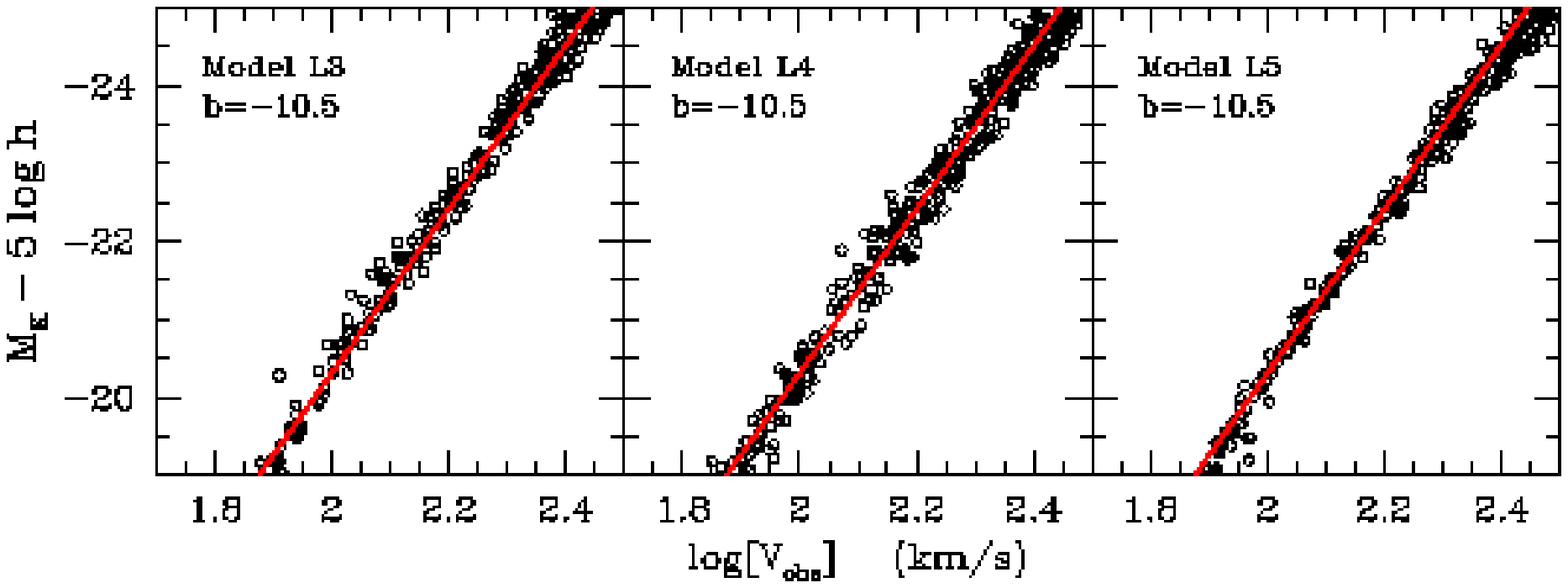}}
\ifsubmode
\vskip3.0truecm
\addtocounter{figure}{1}
\centerline{Figure~\thefigure}
\else\figcaption{\figcapfb}\fi
\end{figure}


\clearpage
\begin{figure}
\epsfxsize=16.0truecm
\centerline{\epsfbox{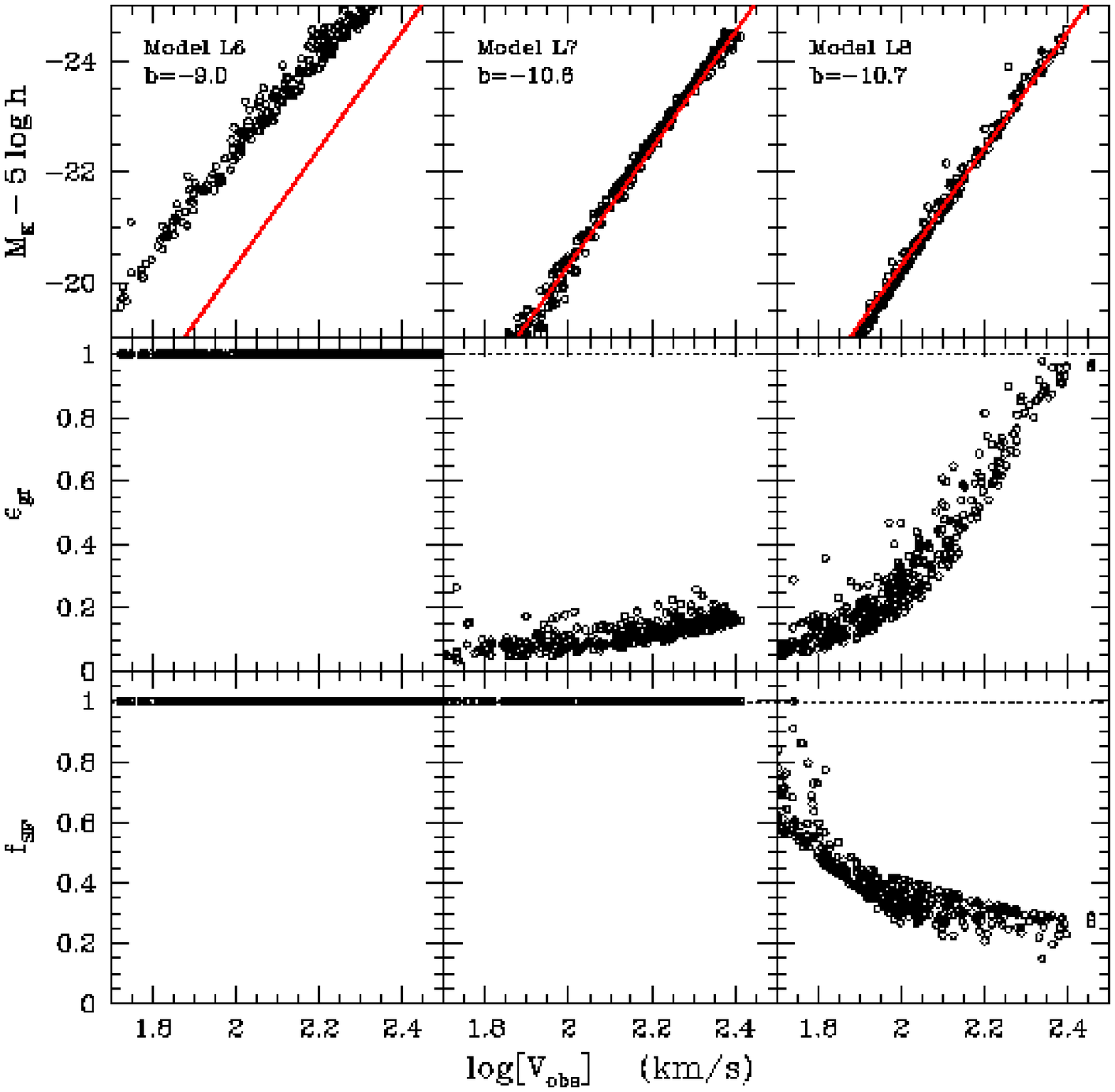}}
\ifsubmode
\vskip3.0truecm
\addtocounter{figure}{1}
\centerline{Figure~\thefigure}
\else\figcaption{\figcapaltencosmo}\fi
\end{figure}


\clearpage
\begin{figure}
\epsfxsize=7.0truecm
\centerline{\epsfbox{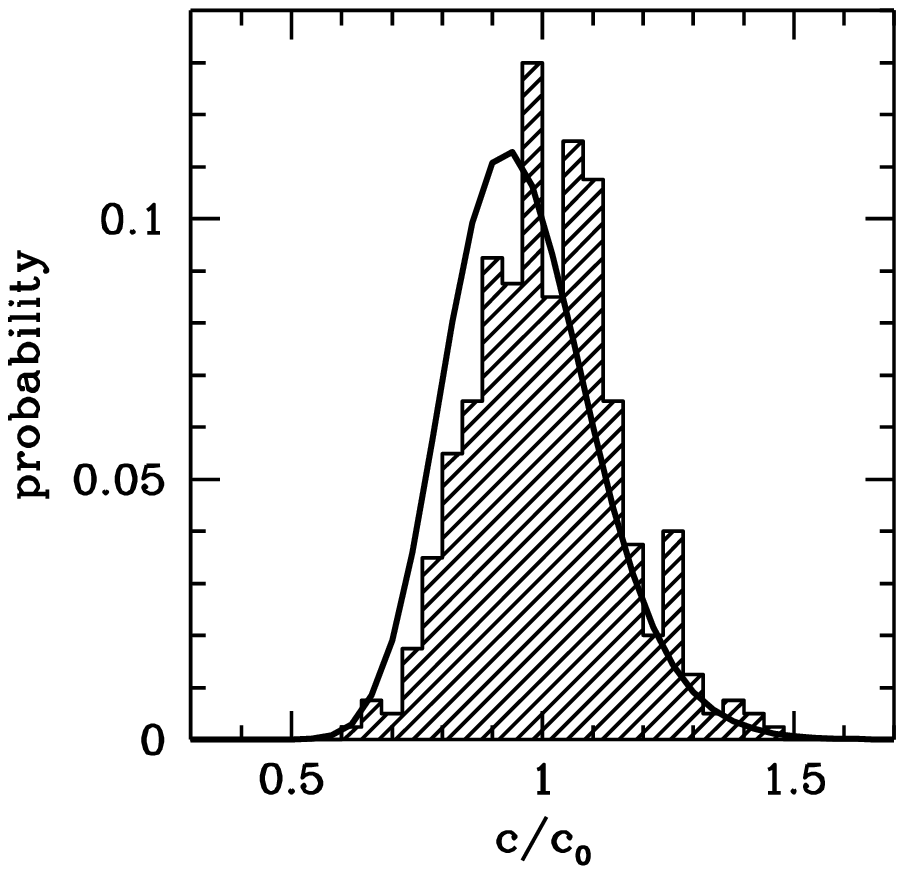}}
\ifsubmode
\vskip3.0truecm
\addtocounter{figure}{1}
\centerline{Figure~\thefigure}
\else\figcaption{\figcapchisto}\fi
\end{figure}


\ifemulate\else
  \clearpage
\fi
\begin{figure}
\epsfxsize=11.0truecm
\centerline{\epsfbox{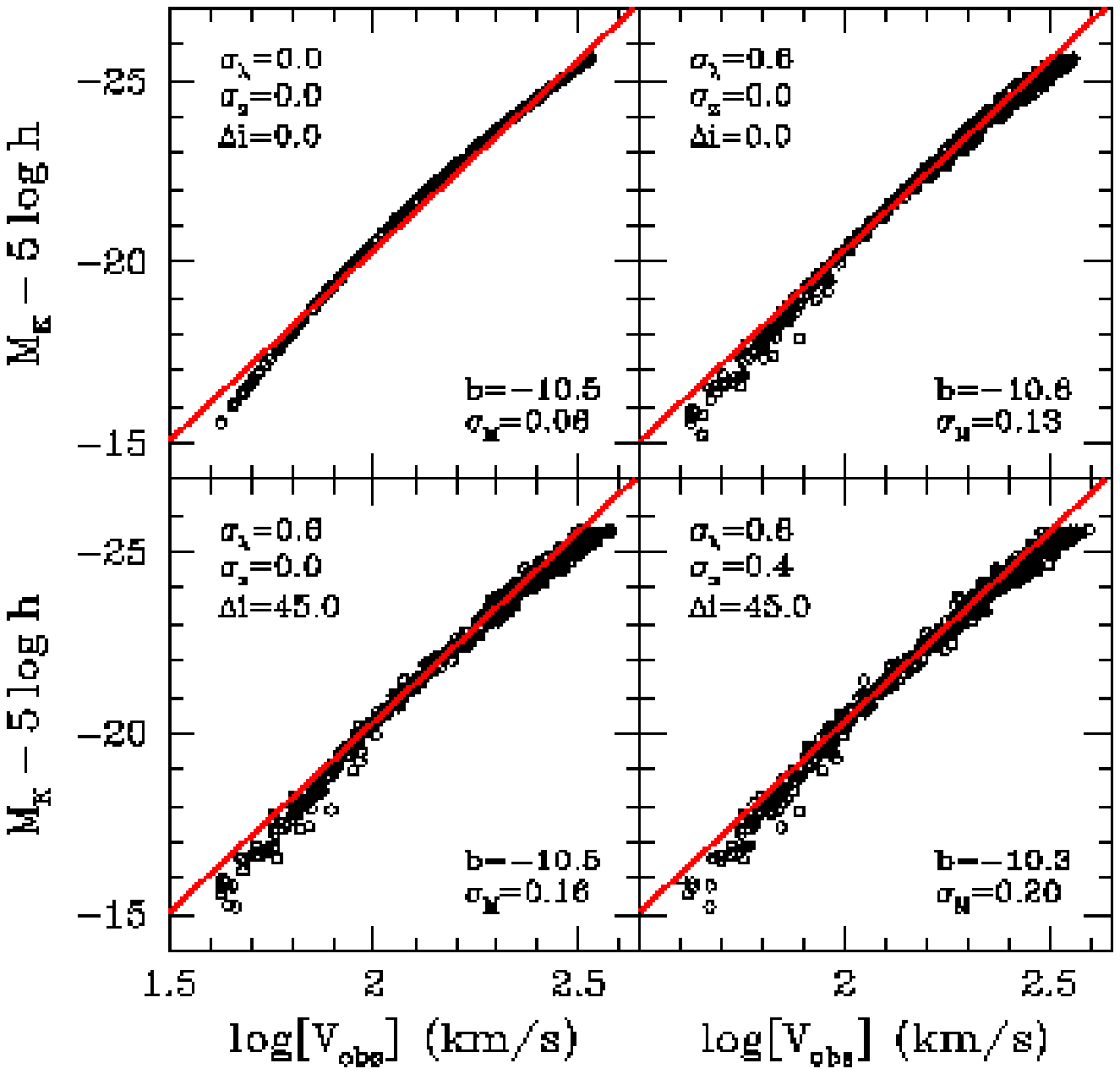}}
\ifsubmode
\vskip3.0truecm
\addtocounter{figure}{1}
\centerline{Figure~\thefigure}
\else\figcaption{\figcapscatter}\fi
\end{figure}


\ifemulate\else
  \clearpage
\fi
\begin{figure}
\epsfxsize=7.0truecm
\centerline{\epsfbox{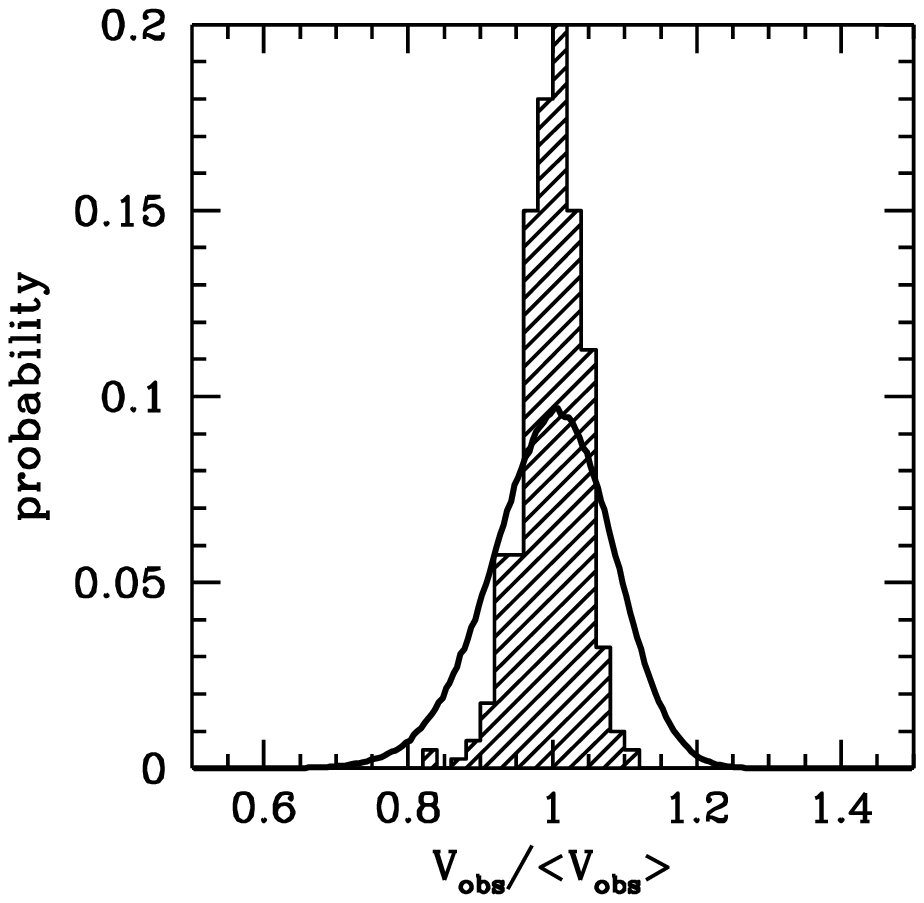}}
\ifsubmode
\vskip3.0truecm
\addtocounter{figure}{1}
\centerline{Figure~\thefigure}
\else\figcaption{\figcapel}\fi
\end{figure}


\clearpage
\begin{figure}
\epsfxsize=10.0truecm
\centerline{\epsfbox{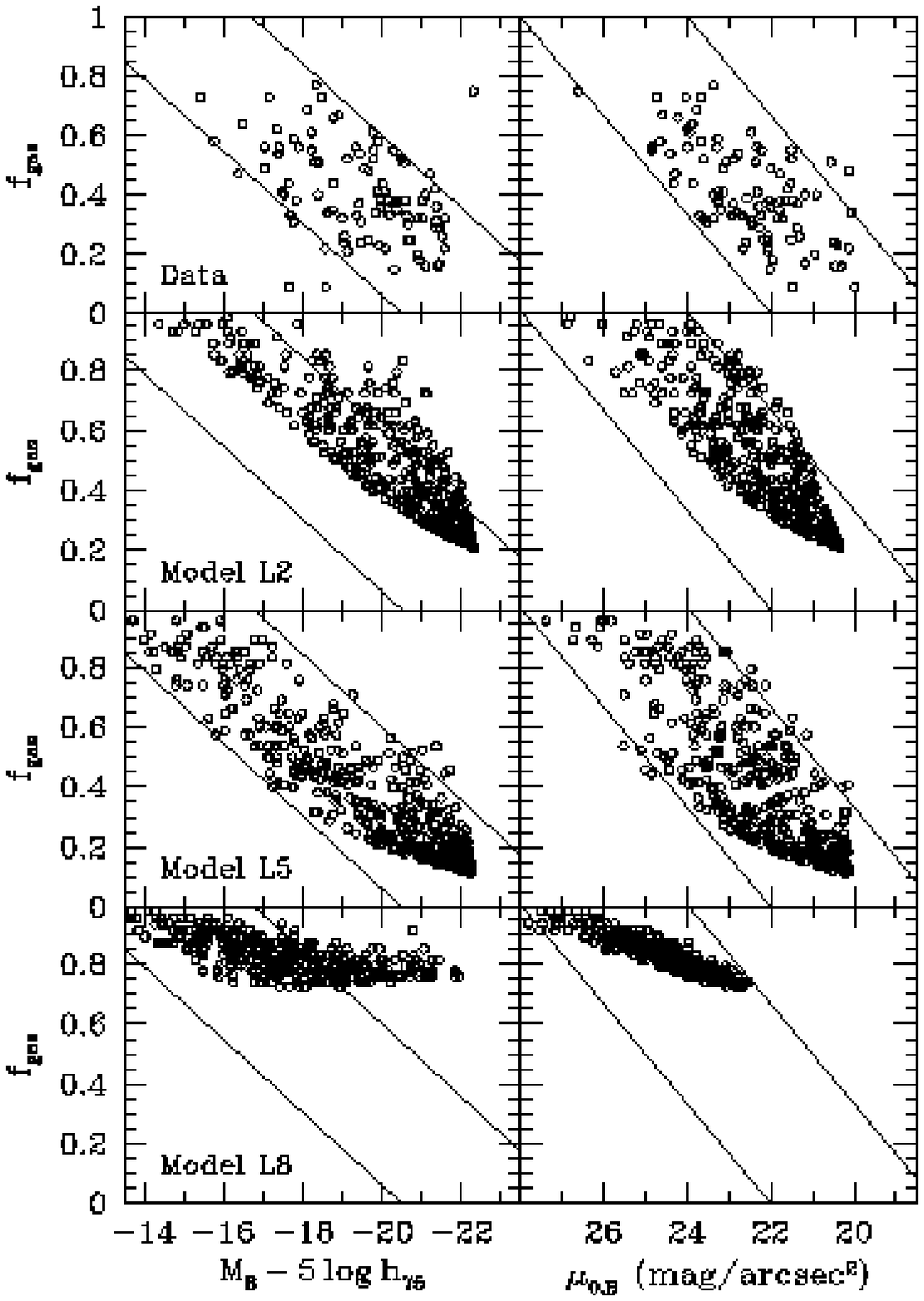}}
\ifsubmode
\vskip3.0truecm
\addtocounter{figure}{1}
\centerline{Figure~\thefigure}
\else\figcaption{\figcapgasfrac}\fi
\end{figure}


\fi



\clearpage
\ifsubmode\pagestyle{empty}\fi


\begin{deluxetable}{lcccllccccc}
\tablecaption{Parameters of cosmological models.\label{tab:cosmo}}
\tablehead{
\colhead{Model} & \colhead{$\Omega_0$} & \colhead{$\Omega_{\Lambda}$} & 
\colhead{$h$} & \colhead{$\sigma_8$} & \colhead{$f_{\rm bar}$} &
\colhead{$t_0$} & \colhead{Clusters} & \colhead{COBE} & 
\colhead{RCs} & \colhead{TF} \\ 
\colhead{(1)} & \colhead{(2)} & \colhead{(3)} & 
\colhead{(4)} & \colhead{(5)} & \colhead{(6)} &
\colhead{(7)} & \colhead{(8)} & \colhead{(9)} & 
\colhead{(10)} & \colhead{(11)}\\
}
\startdata
SCDM & $1.0$ & $0.0$ & $0.5$ & $0.6$ & $0.05$ & $13.0$ & Y & N & N & N \\
$\Lambda$CDM3 & $0.3$ & $0.7$ & $0.7$ & $1.0$ & $0.085$ & $13.5$ & Y & Y & N & Y \\
$\Lambda$CDM2 & $0.2$ & $0.8$ & $0.5$ & $0.38$ & $0.25$ & $21.0$ & N & Y & Y & Y \\
\enddata

\tablecomments{Column~(1) lists the name of the CDM cosmology by which
  we refer to it in the text.  Columns~(2), (3), (4), and (5) list the
  matter    density,  $\Omega_0$, the  density   in    the  form of  a
  cosmological constant, $\Omega_{\Lambda} = \Lambda  / (3 \,  H_0^2)$
  (with $\Lambda$ the cosmological constant), the hubble parameter, $h
  = H_0/(100 \kmsmpc)$,  and the rms mass  variance  on a scale  of $8
  h^{-1}$ Mpc,    $\sigma_8$.  Column~(6) lists   the baryon  fraction
  $f_{\rm bar} = \Omega_{\rm bar}/\Omega_0$  based on a baryon density
  of     $\Omega_{\rm  bar} = 0.0125   \,    h^{-2}$   as required  by
  nucleosynthesis constraints  (e.g., Walker \etal  1991).  Column~(7)
  lists the age  of the universe  in Gyr. Columns~(8) to~(11) indicate
  whether the    models  are consistent   with  the   observed cluster
  abundances, the COBE normalization,  the observed rotation curves of
  spirals (RCs; see Navarro 1998), and the TF zero-point, respectively
  (see \S~\ref{sec:cosmo_cons} for details)}
\end{deluxetable}

\ifemulate\else
  \clearpage
\fi


\begin{deluxetable}{llrcclcrcrc}
\tablecaption{Parameters of models discussed in the text. \label{tab:param}}
\tablehead{
\colhead{ID.} & \colhead{Cosmology} & \colhead{$Q$} & 
\colhead{$\Upsilon^{*}_K$} & \colhead{$n$} & \colhead{$A_{\rm SF}$} &
\colhead{$\varepsilon_{\rm SN}^0$} & \colhead{$\nu$} &
\colhead{$\zeta$} & \colhead{$b$} &\colhead{$\sigma_M$} \\ 
\colhead{(1)} & \colhead{(2)} & \colhead{(3)} & 
\colhead{(4)} & \colhead{(5)} & \colhead{(6)} &
\colhead{(7)} & \colhead{(8)} & \colhead{(9)} &
\colhead{(10)} & \colhead{(11)} \\
}
\startdata
S1 & SCDM & $1.5$ & $0.3$ & $1.4$ & $0.25$ & $0.0$ & $0.0$ & $--$ & $-8.9$ & $0.15$ \\
L0 & $\Lambda$CDM3 & $15.0$ & $0.4$ & $1.4$ & $0.25$ & $0.0$ & $0.0$ & $--$ & $-7.5$ & $0.28$ \\
L1 & $\Lambda$CDM3 & $1.5$ & $0.4$ & $1.4$ & $0.25$ & $0.0$ & $0.0$ & $--$ & $-9.1$ & $0.18$ \\
L2 & $\Lambda$CDM3 & $0.8$ & $0.4$ & $1.4$ & $0.25$ & $0.0$ & $0.0$ & $--$ & $-10.0$ & $0.23$ \\
L3 & $\Lambda$CDM3 & $1.5$ & $\propto (V_{200})^{-0.7}$ & $1.4$ & $0.25$ & $0.0$ & $0.0$ & $--$ & $-10.5$ & $0.20$ \\
L4 & $\Lambda$CDM3 & $1.5$ & $0.4$ & $1.4$ & $0.25$ & $0.0$ & $0.0$ & $0.26$ & $-10.5$ & $0.22$ \\
L5 & $\Lambda$CDM3 & $1.5$ & $0.4$ & $1.4$ & $0.25$ & $0.05$ & $-0.3$ & $--$ & $-10.5$ & $0.16$ \\
L6 & $\Lambda$CDM2 & $1.5$ & $0.4$ & $1.4$ & $0.25$ & $0.0$ & $0.0$ & $--$ & $-9.0$ & $0.19$ \\
L7 & $\Lambda$CDM2 & $1.5$ & $0.4$ & $1.4$ & $0.25$ & $2.0$ & $0.0$ & $--$ & $-10.6$ & $0.13$ \\
L8 & $\Lambda$CDM2 & $1.5$ & $0.4$ & $1.4$ & $0.01$ & $0.05$ & $-2.6$ & $--$ & $-10.7$ & $0.13$ \\
\enddata

\tablecomments{Column~(1) lists the model ID,  by which we refer to it
  in the text.  Column~(2)  lists   the cosmological model  used  (see
  Table~\ref{tab:cosmo}).  Columns~(3)  and~(4) list the  value of the
  Toomre parameter, $Q$, and the $K$-band stellar mass-to-light ratio,
  $\Upsilon_K^{*}$, respectively. The   star formation parameters  $n$
  and $A_{\rm SF}$ are  listed in columns~(5)  and (6),  respectively. 
  Columns~(7)  and~(8) list the  feedback parameters $\varepsilon_{\rm
    SN}^0$ and  $\nu$, respectively.  Column  (9) gives  the parameter
  $\zeta$. This parameter is only defined  for model L3; for all other
  models, the mass dependence of $c$ is determined  from the CDM power
  spectrum.    Finally, columns~(10) and~(11) list  the  slope $b$ and
  scatter $\sigma_M$  (in mag) of the TF   relation for model galaxies
  with $-19  \geq M_K - 5 {\rm  log} h \geq  -24$ and with $\sigma_z =
  0$}
\end{deluxetable}

\clearpage


\end{document}